\documentclass[11pt]{article}

\usepackage{amssymb,amsmath,amsthm,epsfig,isolatin1,enumerate,mathrsfs}
\usepackage{epsfig} 

\theoremstyle{plain}
\newtheorem{thm}{Theorem}[section]
\newtheorem{lem}{Lemma}[section]
\newtheorem{cor}{Corollary}[section]
\newtheorem{prop}{Proposition}[section]

\theoremstyle{definition}

\theoremstyle{remark}

\newtheorem{case}{Case}
\newtheorem{pt}{Part}
\newtheorem{stm}{Statement}

\newcommand{\PR}[1]{\ensuremath{{\mathbf{Pr}\left[#1\right]}}}
\newcommand{\EXP}[1]{\ensuremath{{\mathbf{E}\left(#1\right)}}}

\newcommand{\real}{\ensuremath {\mathbb R} }
\newcommand{\ent}{\ensuremath {\mathbb Z} }
\newcommand{\nat}{\ensuremath {\mathbb N} }

\newcommand{\remove}[1]{}
\newcommand{\bfm}[1] {\mbox{\boldmath$#1$}}

\newcommand{\st}{\;:\;}

\newcommand{\RG} {\ensuremath{\mathcal G(n,r)}}
\newcommand{\UT} {\ensuremath{[0,1)^2}}
\newcommand{\UTT} {\ensuremath{[0,1)^3}}

\newcommand{\nset} {\ensuremath{\{1,\ldots,n\}}}

\newcommand{\nRG} {\ensuremath{G(\cX;r)}}
\newcommand{\RGt} {\ensuremath{G(\cX_t;r)}}
\newcommand{\RGtt} {\ensuremath{G(\cX_{t+1};r)}}
\newcommand{\RGT} {\ensuremath{\big(G(\cX_t;r)\big)_{t\in\ent}}}

\newcommand{\EPi} {\cE_{P,i_1,\ldots,i_k}}

\newcommand{\ar}[1] {\mathsf{Area}(#1)}
\newcommand{\vol}[1] {\mathsf{Vol}(#1)}
\newcommand{\ex} {{\bf E}}
\newcommand{\pr} {{\bf P}}

\newcommand{\tK} {\ensuremath{\widetilde K}}
\newcommand{\tX} {\ensuremath{\widetilde X}}

\newcommand{\cA} {\ensuremath{\mathcal A}}
\newcommand{\cB} {\ensuremath{\mathcal B}}
\newcommand{\cC} {\ensuremath{\mathcal C}}
\newcommand{\cD} {\ensuremath{\mathcal D}}
\newcommand{\cE} {\ensuremath{\mathcal E}}
\newcommand{\cF} {\ensuremath{\mathcal F}}
\newcommand{\cH} {\ensuremath{\mathcal H}}
\newcommand{\cR} {\ensuremath{\mathcal R}}
\newcommand{\cS} {\ensuremath{\mathcal S}}
\newcommand{\cU} {\ensuremath{\mathcal U}}
\newcommand{\cX} {\ensuremath{\mathcal X}}
\newcommand{\cY} {\ensuremath{\mathcal Y}}
\newcommand{\hP} {\ensuremath{\widehat P}}
\newcommand{\hX} {\ensuremath{\widehat X}}
\newcommand{\hcQ} {\ensuremath{\widehat{\mathcal Q}}}
\newcommand{\hcR} {\ensuremath{\widehat{\mathcal R}}}
\newcommand{\hcS} {\ensuremath{\widehat{\mathcal S}}}
\newcommand{\hcP} {\ensuremath{\widehat{\mathcal P}}}
\newcommand{\hcX} {\ensuremath{\widehat{\mathcal X}}}
\newcommand{\hcY} {\ensuremath{\widehat{\mathcal Y}}}

\title{On the Connectivity of Dynamic Random Geometric Graphs\thanks{Partially
supported by the  and the Spanish CYCIT: TIN2004-07925-C03-01
(GRAMMARS). The first and third author are partially supported by 6th Framework under contract 001907 (DELIS). The first author was also supported by \emph{La distinci\'{o} per a
la promoci\'{o} de la recerca de la Generalitat de Catalunya, 2002}.}}

\author{J.~D\' \i az$^1$ \qquad D.~Mitsche$^2$ \qquad X.~P\' erez$^1$ \smallskip \\
{\small
$^1$Llenguatges i Sistemes Inform\`{a}tics, UPC, 08034
Barcelona }\\
{\small $^2$Institut f\"ur Theoretische Informatik, ETH Z\"urich, 8092 Z\"urich} \\
{\small\tt \{diaz,xperez\}@lsi.upc.edu, dmitsche@inf.ethz.ch}}

\date{\today}

\begin{document}
\maketitle
\begin{abstract}
We provide the first analytical results  for the
connectivity of dynamic random geometric graphs - a model of
mobile wireless networks in which vertices move in random (and periodically updated) directions, 
and an edge exists between two vertices if their Euclidean distance is below a given threshold. 
We provide precise asymptotic results for the expected
length of the connectivity and disconnectivity periods of the
network. We believe the formal tools developed in this work could
be of use in future studies in more concrete settings, in the same
manner  as the  development of connectivity threshold for static
random geometric graphs has affected a lot of research done on
ad hoc networks. In the process of proving results for the dynamic
case we also obtain asymptotically precise bounds for the
probability of the existence of a component of fixed size $\ell$,
$\ell \geq 2$, for the static case.
\end{abstract}
\section{Introduction}\label{sec:intro}
{\em Random Geometric graphs} (RGG) have been a very influential and well-studied model of large 
networks, such as sensor networks, where the network agents are represented by the
vertices of the RGG, and the direct connectivity between agents is represented by the edges
(see for example the recent books
\cite{Hekmat06,PlarreKuman}).
Informally, given a radius $r$, a random geometric graph  results
from placing a set of $n$ vertices (agents) uniformly and
independently at random on the unit torus  and connecting two
vertices if and only if their {\em distance} is at most $r$.

In the late 90s, Penrose \cite{Penrose97}, Gupta-Kumar
\cite{GuptaKumar99} and Apple and Russo \cite{AppleRusso02}, gave
accurate estimations for the value of $r$ at which with high
probability, a RGG becomes connected (see Section~2). We denote this value of $r$ by $r_c$. 
Thereafter, hundreds of
researchers have used those basic results on connectivity  to
design algorithms for more efficient coverage, communication and
energy savings in ad hoc networks, and in
particular for sensor networks (see the previously mentioned
books). On the other hand, much
work has been done on the graph theoretic properties of {\em
static} RGG, comprehensively summarized in the monograph of 
M.~D.~Penrose~\cite{Penrose03}. In Section~\ref{sec:static}, we
prove a result on static random geometric graphs, which was not known before (Theorem~1): At the
threshold of connectivity $r_c$ and for any fixed $\ell>1$, the
probability of having some component of size at least $\ell$ other
than the giant component is asymptotically $\Theta
(1/\log^{\ell-1} n)$. Moreover, the most common of such components
are cliques with exact size $\ell$. This result plays an important
role in the derivation of the main result for the dynamic setting, which is explained below.

Recently, there has been an increasing interest for  MANETs
(mobile ad hoc networks). Several models of mobility  have been
proposed in the literature - for an excellent survey of those models we
refer to~\cite{JBAS}.
In all these models, the connections in the network are created and destroyed
as the vertices move closer together or further apart. 
In all previous work,  the authors performed {\bf empirical
studies} on network topology and routing performance. The
paper~\cite{GHSZ} also deals with the problem of maintaining
connectivity  of mobile vertices communicating by radio,
but from an orthogonal perspective to the one in the present
paper - it describes a {\em kinetic data structure} to maintain the
connected components of the union of unit-radius disks moving in
the plane.

 The particular mobility model we are using here (in the literature it is often called
the \emph{Random Walk} model) was introduced 
by Guerin~\cite{Guerin}, and it can be seen as the foundation for most of the mobility models developed afterwards~\cite{JBAS}. 
In the Random Walk model, each vertex selects uniformly at random a direction (angle) in which to travel. The vertices
select their velocities from a given distribution of velocities, and then each vertex moves in its selected
direction at its selected velocity. After some randomly chosen period of time, each vertex halts, selects a new
direction and velocity, and the process repeats. 
An {\bf experimental study} of the 
connectivity of the resulting ad hoc network for different values of $n$ and $r$ for this particular model is presented 
in~\cite{Santi}. As is stated
in the same paper, in many applications which are not life-critical connectivity is an important parameter: 
\emph{"Temporary network disconnections can be 
tolerated, especially if this goes along with a significant decrease of energy consumption."}
In the present paper, we perform the first {\bf analytical study}
of connectivity in the Random Walk model. The particular variant of the model, that we study, is
the following: Given
an initial RGG with $n$ vertices and a radius set to be at the
known connectivity threshold $r_c$ (see Section~\ref{sec:static}),
each vertex chooses independently and uniformly at random an angle
$\alpha \in [0, 2\pi)$, and moves a distance $s$ in that direction
for a period of $m$ steps. Then a new angle is selected
independently for each vertex, and the process repeats. We denote
this graph model the {\em Dynamic Random Geometric Graph}.

Our main result (Theorem~\ref{thm:main}) provides precise asymptotic results
for the expected number of steps that the graph remains connected
once it becomes connected, and the expected number of steps the
graph remains disconnected once it becomes disconnected. These
expressions are given as a function of $n$ and $s$.
Surprisingly, the final expression on the length of connectivity 
does not depend on the size of the intervals between changes of 
angles (as long as the angles do change, after some possibly large number of steps). 
It is worth to note here that the evolution of
connectivity  in the dynamical setting described in
Section~\ref{sec:dynamic} is {\em not} Markovian, in the sense
that staying connected for a large number of steps does have an impact on 
the probability of being connected at the next step. However, one key (and rather
counterintuitive) fact is that, despite this absence of the Markovian property, 
our results, by Lemma~\ref{lem:2step}, depend only on 
the probability of connectivity change in two
consecutive steps.

Throughout the paper, we assume the usual Euclidean distance on the 2-dimensional torus $[0,1)^2$, but
similar results can be obtained for any $\ell_p$-norm, $1\leq
p\leq \infty$. Moreover, our results also can be extended to any
cube $[0,1)^d$ for any $d=\Theta(1)$. 
To the best of
our knowledge, the present work is the first work
in which the dynamic connectivity of RGG is
studied  theoretically. In \cite{DiazAl04} the loosely related
problem of the connectivity of the ad hoc graph produced by $w$ vertices moving randomly
along the edges of a $n\times n$ grid is studied.
In \cite{NainAll} the authors use a similar model to the
one used in the present paper to prove that if the vertices are initially distributed uniformly at random, the distribution remains uniform at any time.
Unless otherwise stated, all our stated results are
asymptotic as $n\to\infty$. As usual, the
abbreviation a.a.s.\ stands for
{\em asymptotically almost surely}, i.e.\ with
probability $1-o(1)$. All logarithms in this paper are natural logarithms.

\section{Static Properties}
\label{sec:static}
In this section, we give  the basic known results about
connectivity of static RGG,  and provide a new bound on the
probability of the existence of components of size $i$, $i \geq 2$, at the
connectivity threshold of a RGG. The threshold for
connectivity of RGG has a long and exciting history, but due to lack
of space, we only refer to \cite{Penrose03}. The formal definition
of random geometric graph is the following: Given a set of $n$
vertices and a positive real $r=r(n)$, each vertex is placed at
some random position in the unit torus $[0,1)^2$ selected
independently and uniformly at random (u.a.r.). We denote by
$X_i=(x_i,y_i)$ the random position of vertex $i$ for $i\in\nset$,
and let $\cX=\cX(n)=\bigcup_{i=1}^n X_i$. We note that with
probability $1$ no two vertices choose the same position and thus
we restrict the attention to the case that $|\cX|=n$. We define
$\nRG$ as the random graph having $\cX$ as the vertex set, and
with an edge connecting each pair of vertices $X_i$ and $X_j$ in
$\cX$ at distance $d(X_i,X_j)\le r$, where $d(\cdot,\cdot)$
denotes the Euclidean distance in the torus.

Let $K_1$ be the random variable counting the number of isolated vertices in $\nRG$. Then, by multiplying the probability that one vertex is isolated by the number of vertices we obtain
$\EXP{K_1} = n (1-\pi r^2)^{n-1} = n e^{-\pi r^2 n} - O(r^4 n)$.

Defining $\mu=n e^{-\pi r^2 n}$, we get that
the asymptotic behaviour of
$\mu$ characterizes the connectivity of $\nRG$:

\par 

\begin{thm}\label{thm:static}\hspace{0cm}
\begin{itemize}
	\item If $\mu\to0$, then a.a.s.\ $\nRG$ is connected.
	\item If $\mu=\Theta(1)$, then a.a.s.\ $\nRG$ consists of one giant component of size $>n/2$ and a Poisson number (with parameter $\mu$) of isolated vertices.
	\item If $\mu\to\infty$, then a.a.s.\ $\nRG$ is disconnected.
\end{itemize}
\end{thm}
{From} the definition of $\mu$ we deduce that $\mu=\Theta(1)$ iff $r=\sqrt{\frac{\log n \pm O(1)}{\pi n}}$. Therefore as a weaker consequence we conclude that
the property of connectivity of $\nRG$ exhibits a sharp threshold at $r=\sqrt{\frac{\log n}{\pi n}}$.
Theorem~\ref{thm:static} also implies that, if $\mu=\Theta(1)$, the components of size $1$ (i.e. isolated vertices) are predominant and have the main contribution to the connectivity of $\nRG$. In fact if $\cC$ (respectively $\cD$) denotes the event that $\nRG$ is connected (respectively disconnected), we have the following
\begin{cor}\label{cor:static}
Assume that $\mu=\Theta(1)$. Then
\[
\pr(\cC) \sim \pr(K_1=0) \sim e^{-\mu}, \qquad \pr(\cD) \sim \pr(K_1>0) \sim 1-e^{-\mu}.
\]
\end{cor}

Therefore, if $\mu=\Theta(1)$, the probability that $\nRG$ has some component of 
size greater than $1$ other than the giant component is $o(1)$. The goal of the remainder of this section 
is to give more precise bounds on this probability. We need a few definitions.

Given a component $\Gamma$ of $\nRG$, $\Gamma$ is {\em embeddable} \remove{if it is contained 
in some square with sides parallel to the axes of the torus and length $1-2r$.
In other words, $\Gamma$ is embeddable}
if it can be mapped into the square $[r,1-r]^2$ by a translation in the torus. 
Embeddable components do not wrap around the torus. 
\remove{Throughout the chapter and often without explicitly mentioning it, we assume in
all geometrical descriptions involving an embeddable component $\Gamma$ that $\Gamma$ is contained in $[r,1-r]^2$ and regard the torus $[0,1)^2$ as the unit square and $d(\cdot,\cdot)$ as the usual Euclidean distance.
Hence terms as ``left'', ``right'', ``above'' and ``below'' are globally defined.
On the other hand,}
Components which are not embeddable must have a large size (at least $\Omega(1/r)$).
Sometimes several non-embeddable components can coexist together. However, 
there are some non-embeddable components which are so spread around the torus that do not allow any 
room for other non-embeddable ones. Call these components {\em solitary}. Clearly, by definition we can have at most one solitary component. We cannot disprove the existence of a solitary component, 
since with probability $1-o(1)$ there exists a giant component of this nature. For components which are not solitary, we 
give asymptotic bounds on the probability of their existence according to their size.

For a fixed integer $\ell\ge1$, let $K_\ell$ be the number of components in $\nRG$ of size exactly $\ell$. 
For any fixed $\epsilon>0$, let $K'_{\epsilon,\ell}$ be the number of components of size  exactly $\ell$ which have all their vertices at distance at most $\epsilon r$ from their leftmost one.
Let $\tK_\ell$ denote the number of components of size  $\geq \ell$ and which are not solitary.

Notice that $K'_{\epsilon,\ell}\le K_\ell\le\tK_\ell$. In the next
theorem, we show that asymptotically all the weight in the
probability that $\tK_\ell>0$ comes from components which also
contribute to $K'_{\epsilon,\ell}$ for $\epsilon$ arbitrarily
small. This implies that at $r_c$, the more common components of
size $\geq\ell$ are cliques of size exactly $\ell$, with all their
vertices close together. 

\begin{lem}\label{lem:EZei}
Let  $\ell\ge2$ be a fixed integer, and $0<\epsilon<1/2$ also fixed. Assume that $\mu=\Theta(1)$. Then,
\[
\ex K'_{\epsilon,\ell}  = \Theta(1/\log^{\ell-1} n)
\]
\end{lem}
\begin{proof}
First observe that with probability $1$, for each component $\Gamma$ which contributes to $K'_{\epsilon,\ell}$, $\Gamma$ has a unique leftmost vertex $X_i$ and the vertex $X_j$ in $\Gamma$ at greatest distance from $X_i$ is also unique. Hence, we can restrict our attention to this case.

Fix an arbitrary set of indices $J\subset\nset$ of size $|J|=\ell$, with two distinguished elements $i$ and $j$.
Denote by $\cY = \bigcup_{k\in J} X_k$ the set of random points in $\cX$ with indices in $J$.
Let $\cE$ be the following event: 
All vertices in $\cY$ are at distance at most $\epsilon r$ from $X_i$ and to the right of $X_i$; vertex $X_j$ is the one in $\cY$ with greatest distance from $X_i$; and the vertices of $\cY$ form a component of $\nRG$.
If $\pr(\cE)$ is multiplied by the number of possible choices of $i$, $j$ and the remaining $\ell-2$ elements of $J$, we get
\begin{equation}\label{eq:EZei}
\ex K'_{\epsilon,\ell} = n(n-1)\binom{n-2}{\ell-2} \pr(\cE).
\end{equation}

In order to bound the probability of $\cE$ we need some definitions.
Let $\rho=d(X_i,X_j)$ and let $\cS$ be the set of all points in the torus $[0,1)^2$ which are at distance at most $r$ from some vertex in $\cY$.
(Notice that $\rho$ and $\cS$ depend on the set of random points $\cY$).
We first need bounds of $\ar{\cS}$ in terms of $\rho$.
Observe that $\cS$ is contained in the circle of radius $r+\rho$ and center $X_i$, and then
\begin{equation}\label{eq:Subound}
\ar{\cS}\le\pi(r+\rho)^2.
\end{equation}
Now let $i_{\mathsf L}=i$, $i_{\mathsf R}$, $i_{\mathsf T}$ and $i_{\mathsf B}$ be respectively the indices of the leftmost, rightmost, topmost and  bottommost vertices in $\cY$ (some of these indices possibly equal).
Assume w.l.o.g. that the vertical length of $\cY$ (i.e.\ the vertical distance between $X_{i_{\mathsf T}}$ and $X_{i_{\mathsf B}}$) is at least $\rho/\sqrt2$. Otherwise, the horizontal length of $\cY$ has this property and we can rotate the descriptions in the argument. The upper halfcircle with center $X_{i_{\mathsf T}}$ and the lower halfcircle with center $X_{i_{\mathsf B}}$ are disjoint and are contained in $\cS$. 
If $X_{i_{\mathsf R}}$ is at greater vertical distance from $X_{i_{\mathsf T}}$ than from $X_{i_{\mathsf B}}$, then consider the rectangle of height $\rho/(2\sqrt2)$ and width $r-\rho/(2\sqrt2)$ with one corner on $X_{i_{\mathsf R}}$ and above and to the right of $X_{i_{\mathsf R}}$. Otherwise, consider the same rectangle below and to the right of $X_{i_{\mathsf R}}$.
This rectangle is also contained in $\cS$ and its interior does not intersect the previously described halfcircles. Analogously, we can find another rectangle of height $\rho/(2\sqrt2)$ and width $r-\rho/(2\sqrt2)$ to the left of $X_{i_{\mathsf L}}$ and either above or below $X_{i_{\mathsf L}}$ with the same properties. Hence,
\begin{equation}\label{eq:Slbound}
\ar{\cS}\ge \pi r^2 + 2\left(\frac\rho{2\sqrt2}\right)\left(r-\frac\rho{2\sqrt2}\right).
\end{equation}
{From}~\eqref{eq:Subound},~\eqref{eq:Slbound} and the fact that $\rho<r/2$, we can write
\begin{equation}\label{eq:Sbound}
\pi r^2 \left(1 + \frac{1}{6} \frac\rho{r} \right) < \ar{\cS} < \pi r^2 \left(1+\frac52\frac\rho r\right) < \frac{9\pi}{4} r^2.
\end{equation}
Now consider the probability $P$ that the $n-\ell$ vertices not in $\cY$ lie outside $\cS$. Clearly $P=(1-\ar{\cS})^{n-\ell}$. Moreover, by~\eqref{eq:Sbound} and using the fact that $e^{-x-x^2} \le 1-x \le e^{-x}$ for all $x\in[0,1/2]$, we obtain
\[
e^{-(1+5\rho/(2 r))\pi r^2n - (9\pi r^2/4)^2n} < P <   \frac{e^{-(1+\rho/(6r))\pi r^2n}}{(1-9\pi r^2/4)^\ell},
\]
and after a few manipulations
\begin{equation} \label{eq:Pbound}
\left(\frac\mu n\right)^{1+5\rho/(2 r)} e^{-(9\pi r^2/4)^2n} < P <  \left(\frac\mu n\right)^{1+\rho/(6r)}    \frac{1}{(1-9\pi r^2/4)^\ell}.
\end{equation}

Event $\cE$ can also be described as follows: There is some nonnegative real $\rho\le\epsilon r$ such that $X_j$ is placed at distance $\rho$ from $X_i$ and to the right of $X_i$; all the remaining vertices in $\cY$ are inside the halfcircle of center $X_i$ and radius $\rho$;
and the $n-\ell$ vertices not in $\cY$ lie outside $\cS$.
Hence, $\pr(\cE)$ can be  bounded from above (below) by integrating with respect to $\rho$ the probability density function of $d(X_i,X_j)$ times the probability that the remaining $\ell-2$ selected vertices lie inside the right halfcircle of center $X_i$ and radius $\rho$ times the upper (lower) bound on $P$ we obtained in~\eqref{eq:Pbound}:
\begin{equation}\label{eq:PE}
\Theta(1)\, I(5/2)\le \pr(\cE) \le \Theta(1)\, I(1/6),
\end{equation}
where
\begin{align}
I(\beta) &= \int_0^{\epsilon r} \pi\rho \left(\frac\pi2\rho^2\right)^{\ell-2} \frac{1}{n^{1 + \beta\rho/r}}  \,d\rho.
\notag\\
&=  \frac2n \left(\frac\pi2 r^2\right)^{\ell-1}  \int_0^{\epsilon} x^{2\ell-3} n^{-\beta x}  \,dx
\label{eq:Ibeta}
\end{align}
Since $\ell$ is fixed, for $\beta=5/2$ or $\beta=1/6$,
\begin{align}
I(\beta) &= \Theta\left(\frac{\log^{\ell-1}n}{n^{\ell}}\right)  \int_0^{\epsilon} x^{2\ell-3} n^{-\beta x}  \,dx
\notag\\
&= \Theta\left(\frac{\log^{\ell-1}n}{n^{\ell}}\right)  \frac{(2\ell-3)!}{(\beta\log n)^{2\ell-2}}
\notag\\
&= \Theta\left(\frac{1}{n^{\ell}\log^{\ell-1}n}\right)
\label{eq:Ibeta2},
\end{align}
and the statement follows from~\eqref{eq:EZei}, \eqref{eq:PE} and~\eqref{eq:Ibeta2}.
\end{proof}
%
%
%
\begin{lem}\label{lem:PY}
Let  $\ell\ge2$ be a fixed integer. Let $\epsilon>0$ be also fixed. Assume that $\mu=\Theta(1)$. Then
\[
\pr(\tK_\ell-K'_{\epsilon,\ell}>0) = O(1/\log^{\ell} n)
\]
\end{lem}
\begin{proof}
We assume throughout this proof that $\epsilon\le10^{-18}$, and prove the claim for this case.
The case $\epsilon>10^{-18}$ follows from the fact that
$(\tK_\ell-K'_{\epsilon,\ell}) \le (\tK_\ell-K'_{10^{-18},\ell})$.

Consider all the possible components in $\nRG$ which are not solitary. Remove from these components the ones of size at most $\ell$ and diameter at most $\epsilon r$, and denote by $M$ the number of remaining components. By construction $\tK_\ell-K'_{\epsilon,\ell} \le M$, and therefore it is sufficient to prove that $\pr(M>0)=O(1/\log^{\ell} n)$.
The components counted by $M$ are classified into several types according to their size and diameter. We deal with each type separately.
\setcounter{pt}{0}
\begin{pt}\label{p:small}
Consider all the possible components in $\nRG$ which have diameter at most $\epsilon r$ and size between $\ell+1$ and  $\log n/37$. Call them components of type~\ref{p:small}, and let $M_{\ref{p:small}}$ denote their number.

For each $k$, $\ell+1\le k\le\log n/37$, let $E_k$ be the expected number of components of type~\ref{p:small} and size $k$.
We observe that these components have all of their vertices at distance at most $\epsilon r$ from the leftmost one. Therefore, we can apply the
same argument we used for bounding $\ex K'_{\epsilon,\ell}$ in the proof of Lemma~\ref{lem:EZei}. Note that~\eqref{eq:EZei}, \eqref{eq:PE} and~\eqref{eq:Ibeta} are also valid for sizes not fixed but depending on $n$. Thus we obtain
\[
E_k \le O(1) n(n-1)\binom{n-2}{k-2} I(1/6),
\]
where $I(1/6)$ is defined in~\eqref{eq:Ibeta}. We use the fact that $\binom{n-2}{k-2}\le(\frac{ne}{k-2})^{k-2}$ and get
\begin{equation}\label{eq:El}
E_k = O(1)  \log n \left(\frac{e}2  \frac{\log n}{k-2}\right)^{k-2} \int_0^{\epsilon} x^{2k-3} n^{-x/6}  \,dx
\end{equation}
The expression $x^{2k-3} n^{-x/6}$ can be maximized for $x\in\real^+$ by elementary techniques, and we deduce that 
\[
x^{2k-3} n^{-x/6} \le \left(\frac{2k-3}{(e/6)\log n}\right)^{2k-3}.
\]
Then we can bound the integral in~\eqref{eq:El} and get
\begin{align*}
E_k &= O(1)  \log n \left(\frac{e}2  \frac{\log n}{k-2}\right)^{k-2}  \epsilon \left(\frac{2k-3}{(e/6)\log n}\right)^{2k-3}
\\
&= O(1)   \left(\frac{36}{2e}  \frac{(2k-3)^2}{(k-2) \log n}\right)^{k-2} k
\end{align*}
Note that for $k\le\log n/37$ the expression $\left(\frac{36}{2e}  \frac{(2k-3)^2}{(k-2) \log n}\right)^{k-2} k$ is decreasing with $k$. Hence we can write
\[
E_k = O \left(\frac{1}{\log^{\ell+1} n}\right), \qquad \forall k \;:\; \ell+3\le k\le\frac1{37}\log n.
\]
Moreover the bounds
$E_{\ell+1}=O(1/\log^{\ell}n)$ and $E_{\ell+2}=O(1/\log^{\ell+1}n)$ are obtained from Lemma~\ref{lem:EZei}, and hence
{\small
\[
\ex M_{\ref{p:small}} = \sum_{k=\ell+1}^{\frac1{37}\log n} E_k
= O\left(\frac{1}{\log^{\ell} n}\right) +  O\left(\frac{1}{\log^{\ell+1} n}\right) + \frac{\log n}{37} O\left(\frac{1}{\log^{\ell+1} n}\right)
=  O\left(\frac{1}{\log^{\ell} n}\right),
\]
}
and then $\pr(M_{\ref{p:small}}>0)\le\ex M_{\ref{p:small}} = O(1/\log^{\ell} n).$
\end{pt}
\begin{pt}\label{p:dense}
Consider all the possible components in $\nRG$ which have diameter at most $\epsilon r$ and size greater than $\log n/37$. Call them components of type~\ref{p:dense}, and let $M_{\ref{p:dense}}$ denote their number.

We tessellate the torus with square cells of side $y=\lfloor(\epsilon r)^{-1}\rfloor^{-1}$ ($y\ge\epsilon r$ but also $y\sim\epsilon r$).
We define a box to be a square of side $2y$ consisting of the union of $4$ cells of the tessellation. Consider the set of all possible boxes.
Note that any component  of  type~\ref{p:dense} must be fully contained in some box.

Let us fix a box $b$. Let $W$ be the number of vertices which are deployed inside $b$. Clearly $W$ has a binomial distribution with mean
$\ex W = (2y)^2n \sim (2\epsilon)^2\log n/\pi$. By
setting $\delta=\frac{\log n}{37\ex W} - 1$  and applying Chernoff inequality to $W$, we have
\[
\pr(W>\frac1{37}\log n) = \pr(W>(1+\delta)\ex W) \le \left(\frac{e^\delta}{(1+\delta)^{1+\delta}}\right)^{\ex W}
= n^{ -\frac{(\log(1+\delta) -  \frac\delta{1+\delta})}{37} }.
\]
Note that $\delta \sim \frac{\pi}{148\epsilon^2}-1> e^{79}$, and therefore
\[
\pr(W>\frac1{37}\log n)  < n^{-2.1}.
\]
Then taking a union bound over the set of all $\Theta(r^{-1})=\Theta(n/\log n)$ boxes, the probability that there is some box with more than $\frac1{37}\log n$ vertices is $O(1/(n^{1.1}\log n))$. Then since each component of type~\ref{p:dense} is contained in some box, we have
\[
\pr(M_{\ref{p:dense}}>0)=O(1/(n^{1.1}\log n)).
\]
\end{pt}
\begin{pt}\label{p:large}
Consider all the possible components in $\nRG$ which are embeddable and have diameter at least $\epsilon r$. Call them components of type~\ref{p:large}, and let $M_{\ref{p:large}}$ denote their number.

We tessellate the torus into square cells of side $\alpha r$, for some $\alpha=\alpha(\epsilon)>0$ fixed but sufficiently small. 
Let $\Gamma$ be a component of type~\ref{p:large}.
Let $\cS=\cS_\Gamma$ be the set of all points in the torus $[0,1)^2$ which are at distance at most $r$ from some vertex in $\Gamma$.
Remove from $\cS$ the vertices of $\Gamma$ and the edges (represented by straight segments) and denote by $\cS'$ the outer connected topologic component of the remaining set. By construction, $\cS'$ must contain no vertex in $\cX$.

Now let $i_{\mathsf L}$, $i_{\mathsf R}$, $i_{\mathsf T}$ and $i_{\mathsf B}$ be respectively the indices of the leftmost, rightmost, topmost and  bottommost vertices in $\Gamma$ (some of these indices possibly equal).
Assume w.l.o.g. that the vertical length of $\Gamma$ (i.e.\ the vertical distance between $X_{i_{\mathsf T}}$ and $X_{i_{\mathsf B}}$) is at least $\epsilon r/\sqrt2$. Otherwise, the horizontal length of $\Gamma$ has this property and we can rotate the descriptions in the argument. The upper halfcircle with center $X_{i_{\mathsf T}}$ and the lower halfcircle with center $X_{i_{\mathsf B}}$ are disjoint and are contained in $\cS'$. 
If $X_{i_{\mathsf R}}$ is at greater vertical distance from $X_{i_{\mathsf T}}$ than from $X_{i_{\mathsf B}}$, then consider the rectangle of height $\epsilon r/(2\sqrt2)$ and width $r-\epsilon r/(2\sqrt2)$ with one corner on $X_{i_{\mathsf R}}$ and above and to the right of $X_{i_{\mathsf R}}$. Otherwise, consider the same rectangle below and to the right of $X_{i_{\mathsf R}}$.
This rectangle is also contained in $\cS'$ and its interior does not intersect the previously described halfcircles. Analogously, we can find another rectangle of height $\epsilon r/(2\sqrt2)$ and width $r-\epsilon r/(2\sqrt2)$ to the left of $X_{i_{\mathsf L}}$ and either above or below $X_{i_{\mathsf L}}$ with the same properties. Hence, taking into account that $\epsilon\le10^{-18}$, we have
\begin{equation}\label{eq:S'bound}
\ar{\cS'}\ge \pi r^2 + 2\left(\frac{\epsilon r}{2\sqrt2}\right)\left(r-\frac{\epsilon r}{2\sqrt2}\right) > \left(1+\frac{\epsilon}{5}\right)\pi r^2.
\end{equation}
Let $\cS^*$ be the union of all the cells in the tessellation which are fully contained in $\cS'$. We loose a bit of area compared to $\cS'$. However, if $\alpha$ was chosen small enough, we can guarantee that $\cS^*$ is topologically connected and has area $\ar{\cS^*}\ge(1+\epsilon/6)\pi r^2$. This $\alpha$ can be chosen to be the same for all components of type~\ref{p:large}.

Hence, we showed that the event $(M_{\ref{p:large}}>0)$ implies that some connected union of cells $\cS^*$ of area $\ar{\cS^*}\ge(1+\epsilon/6)\pi r^2$ contains no vertices. By removing some cells from $\cS^*$, we can assume that
$(1+\epsilon/6)\pi r^2 \le \ar{\cS^*} <  (1+\epsilon/6)\pi r^2 + \alpha^2 r^2$.
Let $\cS^*$ be any union of cells with these properties. (Note that there are $\Theta(1/r^2)=\Theta(n/\log n)$ many possible choices for $\cS^*$.) The probability that $\cS^*$ contains no vertices is
\[
(1-\ar{\cS^*})^n \le e^{-(1+\epsilon/6)\pi r^2 n} = \left(\frac\mu n\right)^{1+\epsilon/6}.
\]
Therefore, we can take the union bound over all the $\Theta(n/\log n)$ possible $\cS^*$,
and obtain an upper bound of the probability that there is some component of the type~\ref{p:large}:
\[
\pr(M_{\ref{p:large}}>0) \le \Theta\left(\frac{n}{\log n}\right) \left(\frac\mu n\right)^{1+\epsilon/6}
= \Theta\left(\frac{1}{n^{\epsilon/6}\log n}\right) .
\]
\end{pt}
\begin{pt}\label{p:notembed}
Consider all the possible components in $\nRG$ which are not embeddable but not solitary either.  Call them components of type~\ref{p:notembed}, and let $M_{\ref{p:notembed}}$ denote their number.

We tessellate the torus $[0,1)^2$ into $\Theta(n/\log n)$ small square cells of side length $\alpha r$, where $\alpha>0$ is a sufficiently small positive constant.

Let $\Gamma$ be a component of type~\ref{p:notembed}.
Let $\cS=\cS_\Gamma$ be the set of all points in the torus $[0,1)^2$ which are at distance at most $r$ from some vertex in $\Gamma$.
Remove from $\cS$ the vertices of $\Gamma$ and the edges (represented by straight segments) and denote by $\cS'$ the remaining set. By construction, $\cS'$ must contain no vertex in $\cX$.

Suppose there is a horizontal or a vertical band of width $2r$ in $[0,1)^2$ which does not intersect the component $\Gamma$ (assume w.l.o.g. that it is the topmost horizontal band consisting of all points with the $y$-coordinate in $[1-2r,1)$).
Let us divide the torus into vertical bands of width $2r$. All of them must contain at least one vertex of $\Gamma$, since otherwise $\Gamma$ would be embeddable. Select any $9$ consecutive vertical bands and pick one vertex of $\Gamma$ with maximal $y$-coordinate in each one. For each one of these $9$ vertices, we select the left upper quartercircle centered at the vertex if the vertex is closer to the right side of the band or the right upper quartercircle otherwise. These nine quartercircles we chose are disjoint and must contain no vertices by construction. Moreover, they belong to the same connected component of the set $\cS'$, which we denote by $\cS''$, and which has an area of $\ar{\cS''}\ge(9/4)\pi r^2$.
Let $\cS^*$ be the union of all the cells in the tessellation of the torus which are completely contained in $\cS''$.
We lose a bit of area compared to $\cS''$. However,
as usual, by choosing $\alpha$ small enough we can guarantee that $\cS^*$ is connected and it has an area of $\ar{\cS^*}\ge(11/5)\pi r^2$.
Note that this $\alpha$ can be chosen to be the same for all components $\Gamma$ of this kind.

Suppose otherwise that all horizontal and vertical bands of width $2r$ in $[0,1)^2$ contain at least one 
vertex of $\Gamma$. Since $\Gamma$ is not solitary it must be possible that it coexists with some other 
non-embeddable component $\Gamma'$. Then all vertical bands or all horizontal bands of 
width $2r$ must also contain some vertex of $\Gamma'$ (assume w.l.o.g. the vertical bands do). Let us 
divide the torus into vertical bands of width $2r$. We can find a simple path $\Pi$ with vertices in $\Gamma'$ which passes 
through $11$ consecutive bands. For each one of the $9$ internal bands, pick the uppermost vertex of 
$\Gamma$ in the band below $\Pi$ (in the torus sense). As before each one of these vertices 
contributes with a disjoint quartercircle which must be empty of vertices, and by the same argument 
we obtain a connected union of cells of the tessellation, which we denote by $\cS^*$, with $\ar{\cS^*}\ge(11/5)\pi r^2$ and containing no vertices.

Hence, we showed that the event $(M_{\ref{p:notembed}}>0)$ implies that some connected union of cells $\cS^*$ with $\ar{\cS^*}\ge(11/5)\pi r^2$ contains no vertices.
By repeating the same argument we used for components of type~\ref{p:large} but replacing $(1+\epsilon/6)\pi r^2$ by $(11/5)\pi r^2$, we get
\[
\pr(M_{\ref{p:notembed}}>0) = \Theta\left(\frac{1}{n^{6/5}\log n}\right) .
\]
\end{pt}
\end{proof}
%
%
%
%
%
%
\begin{lem}\label{lem:EZei2}
Let  $\ell\ge2$ be a fixed integer. Let $0<\epsilon<1/2$ be fixed. Assume that $\mu=\Theta(1)$. Then
\[
\ex[K'_{\epsilon,\ell}]_2 = O(1/\log^{2\ell-2} n)
\]
\end{lem}
\begin{proof}
As in the proof of Lemma~\ref{lem:EZei}, we assume that each component $\Gamma$ which contributes to $K'_{\epsilon,\ell}$ has a unique leftmost vertex $X_i$, and the vertex $X_j$ in $\Gamma$ at greatest distance from $X_i$ is also unique. In fact, this happens with probability $1$.

Choose any two disjoint subsets of $\nset$ of size $\ell$ each, namely $J_1$ and $J_2$, with four distinguished elements $i_1,j_1\in J_1$ and $i_2,j_2\in J_2$.
For $k\in\{1,2\}$, denote by $\cY_k = \bigcup_{l\in J_k} X_l$ the set of random points in $\cX$ with indices in $J_k$.
Let $\cE$ be the event that the following conditions hold for both $k=1$ and $k=2$:
All vertices in $\cY_k$ are at distance at most $\epsilon r$ from $X_{i_k}$ and to the right of $X_{i_k}$; vertex $X_{j_k}$ is the one in $\cY_k$ with greatest distance from $X_{i_k}$; and the vertices of $\cY_k$ form a component of $\nRG$.
If $\pr(\cE)$ is multiplied by the number of possible choices of $i_k$, $j_k$ and the remaining vertices of $J_k$, we get
\begin{equation}\label{eq:EZei2}
\ex[K'_{\epsilon,\ell}]_2 = O(n^{2\ell}) \pr(\cE).
\end{equation}

In order to bound the probability of $\cE$ we need some definitions.
For each $k\in\{1,2\}$, let $\rho_k=d(X_{i_k},X_{j_k})$ and let $\cS_k$ be the set of all the points in the torus $[0,1)^2$ which are at distance at most $r$ from some vertex in $\cY_k$. (Obviously $\rho_k$ and $\cS_k$ depend on the set of random points $\cY_k$.) Also define $\cS=\cS_1\cup\cS_2$.

Let $\cF$ be the event that $d(X_{i_1},X_{i_2})>3r$. This holds with probability $1-O(r^2)$. In order  to bound $\pr(\cE\mid\cF)$, we apply a similar approach to the one in the proof of Lemma~\ref{lem:EZei}. In fact,  observe that if $\cF$ holds then $\cS_1\cap\cS_2=\emptyset$.
Therefore in view of~\eqref{eq:Sbound} we can write
\begin{equation}	\label{eq:Sbound2}
\pi r^2(2+(\rho_1+\rho_2)/(6r))<\ar{\cS}<\frac{18\pi}{4}r^2,
\end{equation}
and using the same elementary techniques that gave us~\eqref{eq:Pbound} we get
\begin{equation}\label{eq:Pbound2}
(1-\ar{\cS})^{n-2\ell} <  \left(\frac\mu n\right)^{2+(\rho_1+\rho_2)/(6r)} \frac1{(1-18\pi r^2/4)^{2\ell}}.
\end{equation}
Now observe that $\cE$ can also be described as follows: For each $k\in\{1,2\}$ there is some nonnegative real $\rho_k\le\epsilon r$ such that $X_{j_k}$ is placed at distance $\rho_k$ from $X_{i_k}$ and to the right of $X_{i_k}$; all the remaining vertices in $\cY_k$ are inside the halfcircle of center $X_{i_k}$ and radius $\rho_k$;
and the $n-\ell$ vertices not in $\cY_k$ lie outside $\cS_k$.
In fact, rather than this last condition we only require for our bound that all vertices in $\cX\setminus (\cY_1\cup \cY_2)$ are placed outside $\cS$. Clearly, this has probability $(1-\ar{\cS})^{n-2\ell}$. Then, from~\eqref{eq:Pbound2} and  following an analogous argument to the one that leads to~\eqref{eq:PE},
we obtain the bound
\begin{align*}
\pr(\cE\mid\cF) &\le \Theta(1) \int_0^{\epsilon r}\int_0^{\epsilon r} \pi\rho_1 \left(\frac\pi2 \rho_1^2\right)^{\ell-2} \pi\rho_2 \left(\frac\pi2 \rho_2^2\right)^{\ell-2}
\frac1{n^{2+(\rho_1+\rho_2)/(6r)}} \, d\rho_1d\rho_2
\\
&= \Theta(1) \: I(1/6)^2,
\end{align*}
where $I(1/6)$ is defined in~\eqref{eq:Ibeta}. Thus from~\eqref{eq:Ibeta2} we conclude
\begin{equation}\label{eq:PEF}
\pr(\cE\wedge\cF) \le \Theta(1) \: P(\cF) \: I(1/6)^2 = O\left(\frac{1}{n^{2\ell} \log^{2\ell-2} n}\right).
\end{equation}

Otherwise, suppose that $\cF$ does not hold (i.e.\ $d(X_{i_1},X_{i_2})\le3r$). Observe that $\cE$ implies that $d(X_{i_1},X_{i_2})>r$, since $X_{i_1}$ and $X_{i_2}$ must belong to different components. Hence the circles with  centers on $X_{i_1}$ and $X_{i_2}$ and radius $r$ have an intersection of area less than $(\pi/2)r^2$. These two circles are contained in $\cS$
and then we can write $\ar{\cS}\ge(3/2)\pi r^2$.
Note that $\cE$ implies that all vertices in $\cX\setminus (\cY_1\cup \cY_2)$ are placed outside $\cS$ and that for each $k\in\{1,2\}$ all the vertices in $\cY_k\setminus\{X_{i_k}\}$ are at distance at most $\epsilon r$ and to the right of $X_{i_k}$.
This gives us the following rough bound
\[
\pr(\cE\mid\overline\cF) \le \left(\frac\pi2(\epsilon r)^2\right)^{2\ell-2}  \left(1-\frac{3\pi}2 r^2\right)^{n-2\ell}
= O(1) \left(\frac{\log n}{n}\right)^{2\ell-2}  \left(\frac\mu n\right)^{3/2}.
\]
Multiplying this  by $\pr(\overline\cF)=O(r^2)=O(\log n/n)$ we obtain
\begin{equation}\label{eq:PEnF}
\pr(\cE\wedge\overline\cF) = O\left(\frac{\log^{2\ell-1} n}{ n^{2\ell+1/2}}\right),
\end{equation}
which is negligible compared to \eqref{eq:PEF}. The statement follows from~\eqref{eq:EZei2}, \eqref{eq:PEF} and~\eqref{eq:PEnF}.
\end{proof}

The main result of this section now follows easily.
\begin{thm}\label{thm:static2}
Let  $\ell\ge2$ be a fixed integer.  Let $0<\epsilon<1/2$ be fixed. Assume that $\mu=\Theta(1)$. Then
\[
\PR{\tK_\ell>0} \sim \PR{K_\ell>0} \sim \PR{K'_{\epsilon,\ell}>0}
= \Theta\left(\frac1{\log^{\ell-1} n}\right).
\]
\end{thm}
\begin{proof}
{From} Corollary~1.12 in~\cite{Bollobas01}, we have
\[
\ex K'_{\epsilon,\ell} - \frac12\ex [K'_{\epsilon,\ell}]_2 \le \pr(K'_{\epsilon,\ell}>0) \le \ex K'_{\epsilon,\ell},
\]
and therefore by Lemmata~\ref{lem:EZei} and~\ref{lem:EZei2} we obtain
\[
\pr(K'_{\epsilon,\ell}>0) = \Theta(1/\log^{\ell-1} n).
\]
Combining this and Lemma~\ref{lem:PY}, yields the statement.
\end{proof}

\section{Dynamic Properties}
\label{sec:dynamic}
We define the dynamic model as follows. Given a positive real $s=s(n)$ and a positive integer $m=m(n)$,
we consider the following random process $(\cX_t)_{t\in\ent} = (\cX_t(n,s,m))_{t\in\ent}$:
At time step $t=0$, $n$ agents are scattered independently and u.a.r.\ over the torus $[0,1)^2$, as in the static model.
Moreover each agent chooses u.a.r.\ an
angle $\alpha\in[0,2\pi)$, and moves in the direction of $\alpha$, travelling distance $s$ at each time step. These directions are changed every $m$ steps for all agents. More formally, for each agent $i$ and for each interval $[t,t+m]$ with $t\in\ent$ divisible by $m$, an angle in $[0,2\pi)$ is chosen independently and u.a.r., and this angle determines the direction of $i$ between time steps $t$ and $t+m$. Note that we are also considering negative steps, which is interpreted as if the agents were already moving around the torus ever before step $t=0$.
We extend the notation from the static model, and denote by $X_{i,t}=(x_{i,t},y_{i,t})$ the position of each agent $i$ at time $t$. Also let $\cX_t=\bigcup_{i=1}^n X_{i,t}$ be the set of positions of the agents at time $t$.
Furthermore, given a positive $r=r(n)\in\real$ such that $r=o(1)$, a random graph 
process can be derived from $(\cX_t)_{t\in\ent}$. For any $t\in\ent$, the vertex set of $\RGt$ is $\cX_t$, and 
we join by an edge all pairs of vertices in $\cX_t$ which are at Euclidean distance (in the torus) at most $r$.
We derive asymptotic results on $\RGT$ as $n\to\infty$. 

We use the following lemma proven in~\cite{NainAll}.

\begin{lem}\label{lem:invariant}
At any fixed step $t\in\ent$, the vertices are distributed over the torus \UT\ independently and u.a.r.
Consequently for any $t\in\ent$, $\RGt$ has the same distribution as $\nRG$.
\end{lem}
In the remaining of the section, we focus our attention around the threshold of connectivity and we assume that $\mu=\Theta(1)$, or equivalently
\[
r=\sqrt{\frac{\log n\pm O(1)}{\pi n}}.
\]

In order to prove  the main statement of this section, we first require some technical results which involve only two arbitrary consecutive steps $t$ and $t+1$ of $(\cX_t)_{t\in\ent}$.
In this context $t$ is considered to be an arbitrary fixed integer, and it is often omitted from notation whenever it is understood.
Thus for each $i\in\nset$, the random positions $X_{i,t}$ and $X_{i,t+1}$ of agent $i$ at times $t$ and $t+1$ are simply denoted by $X_i=(x_i,y_i)$ and $X'_i=(x'_i,y'_i)$. Let also $\cX=\cX_t$ and $\cX'=\cX_{t+1}$.
Note that the random points $X_i$ and $X'_i$ are not independent.
In fact if $2\pi z_i$ ($z_i\in[0,1)$) is the angle in which the agent $i$ moves between times $t$ and $t+1$, then $x'_i=x_i+s\cos(2\pi z_i)$ and $y'_i=y_i+s\sin(2\pi z_i)$ (where all the sums involving coordinates are taken$\mod 1$).
This motivates an alternative description of the model at times $t$ and $t+1$ in terms of a three-dimensional placement of the agents, in which the third dimension is interpreted as a normalized angle. For each $i\in\nset$, define the random point $\hX_i=(x_i,y_i,z_i)\in\UTT$, and also let $\hcX=\bigcup_{i=1}^n \hX_i$.
Observe that by Lemma~\ref{lem:invariant} all the random points $\hX_i$ are chosen independently and u.a.r.\ from the $3$-torus \UTT, and also that $\hcX$ encodes all the information of the model at times $t$ and $t+1$.
In fact, if we map \UTT\ onto \UT\ by the following surjections
\begin{align*}
\pi_1 : \UTT &\to \UT & \pi_2 : \UTT &\to \UT \\
(x,y,z) &\mapsto (x,y) & (x,y,z) &\mapsto (x+s\cos(2\pi z),y+s\sin(2\pi z)),
\end{align*}
we can easily recover the positions of agent $i$ at times $t$ and $t+1$ from $\hX_i$ and write $X_i=\pi_1(\hX_i)$ and $X'_i=\pi_2(\hX_i)$.
Notice moreover that, for any measurable set $\cA\subseteq\UT$, the events $X_i\in \cA$ and $X'_i\in \cA$ are respectively equivalent to the events $\hX_i\in\pi_1^{-1}(\cA)$ and $\hX_i\in\pi_2^{-1}(\cA)$ in this new setting.  Furthermore, we have
\begin{equation}\label{eq:AeqV}
\ar\cA =\vol{\pi_1^{-1}(\cA)}=\vol{\pi_2^{-1}(\cA)},
\end{equation}
since
\[
\vol{\pi_1^{-1}(\cA)} = \vol{\cA\times[0,1)} = \ar{\cA},
\]
and also by putting $\cA_z = \cA-(s\cos(2\pi z),s\sin(2\pi z))$
\[
\vol{\pi_2^{-1}(\cA)} = \int_{[0,1)} \Big(\int_{\cA_z} dxdy \Big) \,dz =  \ar{\cA}.
\]
Naturally~\eqref{eq:AeqV} is compatible with the fact that, in view of Lemma~\ref{lem:invariant}, for any measurable sets $\cA\subseteq\UT$ and $\cB\subseteq\UTT$,
\begin{equation*}
\pr(X_i\in \cA) = \ar\cA, \qquad \pr(X'_i\in \cA) = \ar\cA, \qquad \pr(\hX_i\in \cB) = \vol\cB.
\end{equation*}

Now we define some sets which will repeatedly appear in this section.
For each $i\in\nset$, consider the sets
\[
\cR_i = \{X\in\UT \st d(X,X_i)\le r\} \quad\text{and}\quad \cR'_i = \{X\in\UT \st d(X,X'_i)\le r\},
\]
and also let $\hcR_i=\pi_1^{-1}(\cR_i)$ and $\hcR'_i=\pi_2^{-1}(\cR'_i)$ be their counterparts in \UTT.
Note that for $i,j\in\nset$, we have that $\hX_i\in\hcR_j$ iff $d(X_i,X_j)\le r$, and similarly that $\hX_i\in\hcR'_j$ iff $d(X'_i,X'_j)\le r$ (each of these events occurring with probability exactly $\vol{\hcR_i}=\vol{\hcR'_i}=\pi r^2$).
Also observe that $X_i$ is isolated in $G(\cX;r)$ 
iff $(\hcX\setminus\{\hX_i\}) \cap \hcR_i = \emptyset$, and that analogously 
$X'_i$ is isolated in $G(\cX';r)$ iff $(\hcX\setminus\{\hX_i\}) \cap \hcR'_i = \emptyset$.

We need the following
%
%
\begin{lem}\label{lem:tec}
Assume $\mu=\Theta(1)$.
There exists a constant $\epsilon>0$ such that the following statements are true (for large enough $n$): For any $i,j\in\nset$ (possibly $i=j$), 
\begin{enumerate}
\item if $d(X_i,X_j)>r$ then $\vol{\hcR_i\cap\hcR_j} \le \frac\pi2r^2$.
\item if $s<r/7$ and $d(X_i,X_j)>r-2s$ then $\vol{(\hcR_i\cup\hcR'_i)\cap(\hcR_j\cup\hcR'_j)} \le (1-\epsilon)\pi r^2$.
\item if $s\ge r/7$ and $s=O(r)$ then $\vol{\hcR_i\cap\hcR'_j} \le (1-\epsilon)\pi r^2$.
\item if $s=\omega(r)$ then $\vol{\hcR_i\cap\hcR'_j} = O(r^3 \frac{s+1}{s})=o(r^2)$.
\end{enumerate}
\end{lem}
\begin{proof}\hspace{0cm}\par
\setcounter{stm}{0}
\begin{stm}
Assume w.l.o.g. that the segment $\overline{X_iX_j}$ is vertical and that $X_i$ is above $X_j$. Then, let $\cS\subset\UT$ be the upper halfcircle with center $X_i$ and radius $r$, and $\hcS=\pi_1^{-1}(\cS)=\cS\times[0,1)\subset [0,1)^3$. Clearly, $\vol\hcS=\pi r^2/2$, $\hcS\subset\hcR_i$ and $\hcS\cap\hcR_j=\emptyset$, and the statement follows.
\end{stm}

\begin{stm}
The distance between $X'_i$ and $X'_j$ is greater than $3r/7$, since $d(X'_i,X'_j)\ge d(X_i,X_j)-2s>r-4s$. Let $\cS_i$ (respectively $\cS_j$) be the set of points in \UT\ at distance at most $8r/7$ from $X'_i$ (respectively $X'_j$)
Note that $\cS_i$ and $\cS_j$ are two circles of radius $8r/7$ with centers at distance greater than $3r/7$. Then straightforward computations show that $\ar{\cS_i\cap\cS_j}$ is at most $(1-\epsilon)\pi r^2$ for some $\epsilon>0$. We
define $\hcS_i=\pi_1^{-1}(\cS_i)$ and $\hcS_j=\pi_1^{-1}(\cS_j)$. Clearly, $\hcS_i\supset\hcR_i\cup\hcR'_i$ and $\hcS_j\supset\hcR_j\cup\hcR'_j$. Hence,
\[
\vol{(\hcR_i\cup\hcR'_i)\cap(\hcR_j\cup\hcR'_j)} \le \vol{\hcS_i\cap\hcS_j} = \ar{\cS_i\cap\cS_j} \le (1-\epsilon)\pi r^2.
\]
\end{stm}

\begin{stm}
Let $k\in\nset$ be different from $i$ and $j$.
Observe that $\vol{\hcR_i\setminus \hcR'_j}$ is the probability that $d(X_i,X_k)\le r$ but $d(X'_j,X'_k)> r$. Suppose that $d(X_i,X_k)\le r$ but also $d(X'_j,X_k)>13r/14$. (This happens with probability at least $(1-13^2/14^2)\pi r^2$.) Let $\alpha$ be the angle of $\overrightarrow{X'_jX_k}$ with respect to the horizontal axis.
Recall that agent $k$ moves between time steps $t$ and $t+1$ towards a direction $2\pi z_k$, where $z_k$ is the third coordinate of $\hX_k$.
If $2\pi z_k\in[\alpha-\pi/3,\alpha+\pi/3]$, then the agent increases its distance with respect to $X'_j$ by at least $s/2\ge r/14$ and thus $d(X'_j,X'_k)> r/14+13r/14=r$.
This range of directions has probability $1/3$. Summarizing, we proved that $\vol{\hcR_i\setminus\hcR'_j} \ge (1-13^2/14^2)\pi r^2/3$, and the statement follows.
\end{stm}

\begin{stm}
Given $k\in\nset$ different from $i$ and $j$, observe that $\vol{\hcR_i\cap\hcR'_j}$ is the probability that $d(X_k,X_i)\le r$ and also $d(X'_k,X'_j)\le r$.
Suppose first that $s<1/2$.
We claim that the probability that $d(X'_k,X'_j)\le r$ conditional upon any fixed outcome of $X_k$ is at most $(2+\epsilon)r/s$ for some $\epsilon>0$, no matter which particular point $X_k$ is chosen.
In fact, assume $X_k\neq X'_j$ (the case $X_k=X'_j$ is trivial) and let $\alpha$ be the angle of $\overrightarrow{X_kX'_j}$ with respect to the horizontal axis.
If agent $k$ moves between steps $t$ and $t+1$ towards a direction $2\pi z_k$ not in $[\alpha-\arcsin(r/s),\alpha+\arcsin(r/s)]$ then $d(X'_k,X'_j)>r$. Hence, $\vol{\hcR_i\cap\hcR'_j}$ is at most $\pr(d(X_k,X_i)\le r)=\pi r^2$ times $(2+\epsilon)r/s$, which satisfies the statement.
 
The case $s\ge1/2$ is a bit more delicate, since agent $k$ may loop many times around the torus while moving between steps $t$ and $t+1$.
In fact, as we move along the circumference of radius $s$ centered on $X_k$ we cross the axes of the torus $\Theta(1+s)$ times. This gives the extra factor $(1+s)$ in the statement, which is negligible when $s=o(1)$ but grows large when $s=\omega(1)$.\qedhere
\end{stm}
\end{proof}
%
%

For each $i\in\nset$, we define $\hcQ_i =\hcR'_i\setminus\hcR_i$ and
$\hcQ'_i=\hcR_i\setminus\hcR'_i$.
Given any two agents $i$ and $j$, observe that $\hX_i\in\hcQ'_j$ iff $\hX_j\in\hcQ'_i$ iff
$d(X_i,X_j)\le r$ and $d(X'_i,X'_j)>r$ (i.e.\ the agents are joined by an edge at time $t$ but not at time $t+1$). This holds with probability $\vol{\hcQ_i}=\vol{\hcQ'_i}$, which does not depend of the particular agents and of $t$ and will be denoted by $q$ hereinafter. The value of this parameter depends on the asymptotic relation between $r$ and $s$ and is given in the following
%
%
\begin{lem}\label{lem:q}
The probability that two different agents $i,j\in\nset$ are at distance at most $r$ at time $t$ but greater than $r$ at time $t+1$ is $q\le\pi r^2$, which also satisfies
\[
q \sim \begin{cases}
\frac4\pi sr & \text{if }s=o(r),\\
\Theta(r^2) & \text{if }s=\Theta(r),\\
\pi r^2 & \text{if }s=\omega(r).\\
\end{cases}
\]
\end{lem}
\begin{proof}
The first bound on $q$ is immediate from the definition of $q$ and the fact that $\vol{\hcR_i}=\pi r^2$. In order to obtain the second statement, we consider separate cases.
\setcounter{case}{0}
\begin{case}[$s\le\epsilon r$, for some fixed but small enough $\epsilon>0$]
In order to compute the probability that $\hX_j\in\hcQ'_i$, we express $\hX_j=(x_j,y_j,z_j)$ in new coordinates $(\rho,\theta,z)$, where $\rho = d(X_j,X'_i)$, $\theta$ is the angle between the horizontal axis and $\overrightarrow{X_iX_j}$, and $z=z_j$.
Then we integrate an element of volume over the region $\hcQ'_i$ in terms of these coordinates.
Let us also call $\xi = d(X_j,X_i)$, so that $(\xi,\theta,z)$ are the usual cylindrical coordinates. From the law of cosines, we can stablish the relation between $\rho$ and $\xi$ and write
\begin{equation} \label{eq:rhoxi}
\rho = \sqrt{\xi^2 + s^2 - 2 \xi s\cos\theta} \quad\text{and}\quad
\xi = \sqrt{\rho^2 - s^2\sin^2\theta} + s\cos\theta.
\end{equation}
Now observe that the minimum value that $\rho$ can take is $r-s$, since $X_j$ must lie outside the circle of radius $r-s$ and center $X'_i$.
Otherwise by the triangular inequality $d(X'_i,X'_j)\le r$ and the agents $i$ and $j$ would share an edge at step $t+1$.
On the other hand, $X_j$ must lie inside the circle of radius $r$ centered on $X_i$, and therefore (by setting $\xi=r$ in~\eqref{eq:rhoxi}) the maximum value that $\rho$ can achieve is
\[
\rho = \sqrt{r^2 + s^2 - 2 r s\cos\theta}
\]
Moreover, let $\alpha$ be the angle determined from the range of all possible values of $2\pi z$ (i.e.\ possible directions for agent $j$ to move). Again by the law of cosines,
\[
\alpha=2 \arccos\left(\frac{r^2-s^2-\rho^2}{2s\rho}\right).
\]
Finally from~\eqref{eq:rhoxi} and the change of variables formula, it is straighforward to determine the element of volume in coordinates $(\rho,\theta,z)$:
\[
dxdydz = \xi \, d\xi d\theta dz  = \frac{\xi\rho }{\xi-s\cos\theta} \, d\rho d\theta dz,
\]
where, using the fact that $r-2s\le\xi\le r$, we can write
\[
\frac{\xi\rho}{\xi-s\cos\theta} = \rho \left(1 \pm O\left(\frac{s}{r}\right)\right).
\]
In view of all the above, we deduce
\begin{align*}
q&=\int_{\hcQ'_i} dxdydz
\\
&=\int_0^{2\pi} \int_{r-s}^{\sqrt{r^2 + s^2 - 2 r s\cos\theta}} \frac\alpha{2\pi} \frac{\xi\rho}{\xi-s\cos\theta} \,d\rho d\theta
\\
&= \left(1\pm O\left(\frac{s}{r}\right)\right) \int_0^{2\pi} \int_{r-s}^{\sqrt{r^2 - s^2 \sin^2\phi } - s\cos\phi} \frac1{\pi}\arccos\left(\frac{r^2-s^2-\rho^2}{2s\rho}\right) \rho \,d\rho d\theta
\\
&= \left(1\pm O\left(\frac{s}{r}\right)\right) 2 \int_0^{\pi} \frac1{2\pi}
\bigg(-rs\sin\theta   - \theta r^2 \\ &\qquad\qquad\qquad\qquad\qquad
+(r^2+s^2-2rs\cos\theta)\arccos{\frac{r\cos\theta-s}{\sqrt{r^2+s^2-2rs\cos\theta}}}\bigg) d\theta.
\end{align*}
Now by looking at the Taylor series with respect to $s/r$ of the expression inside the integral divided by $r^2$, we get
\begin{align}
q &=  \left(1\pm O\left(\frac{s}{r}\right)\right) \int_0^{\pi} r^2 \left( -\frac{2\theta\cos\theta}{\pi} \frac{s}{r} + O\left(\left(\frac{s}{r}\right)^2\right)\right) d\theta = \left(1\pm O\left(\frac{s}{r}\right)\right) \frac4\pi sr.
\label{eq:q}
\end{align}
\end{case}
\begin{case}[$\epsilon r<s<r/7$]
Recall that $\cR_i$ is the circle of radius $r$ and center $X_i$. 
Take the chord in $\cR_i$ which is perpendicular to the segment $\overline{X_iX'_i}$ and at distance $r$ from $X'_i$. This chord divides $\cR_i$ into two regions. One of them (call it $\cS$) has the property that all the points inside are at distance at least $r$ from $X'_i$ and moreover $\ar\cS \ge \epsilon\sqrt{2\epsilon-\epsilon^2} r^2$.
Suppose that $X_j\in\cS$ (i.e.\ the agent $j$ is in $\cS$ at time $t$), which happens with probability at least $\epsilon\sqrt{2\epsilon-\epsilon^2} r^2$.
Let us now consider the circle centered on $X'_i$ and passing through $X_j$.
We observe that $d(X'_j,X'_i)>d(X_j,X'_i)$ with probability at least $1/2$, since it is sufficient that the direction $2\pi z_j$ in which agent $j$ moves lies in the outer side of the tangent of that circle at $X_j$.
Therefore, the probability that $d(X_j,X_i)\le r$ and $d(X'_j,X'_i)>r$ (i.e. $\hX_j\in\hcQ'_i$) is at least $\frac12\epsilon\sqrt{2\epsilon-\epsilon^2} r^2$.
\end{case}
\begin{case}[$s\ge r/7$]
We can write
\[
q = \vol{\hcQ'_i} = \vol{\hcR_i\setminus\hcR'_i} = \vol{\hcR_i} - \vol{\hcR_i\cap\hcR'_i},
\]
and the result follows from the statements 3--4 in Lemma~\ref{lem:tec}.\qedhere
\end{case}
\end{proof}
%
%
We also need the following
\begin{lem}\label{lem:inclusion_exclusion}
Consider a setting with $n$ balls and $k+1$ disjoint bins
$\cU_0,\ldots,\cU_k$,
where each ball is either placed into at most one of the bins or
possibly remains outside all of them. Suppose that each ball is assigned
to bin $U_i$ with probability $p_i=p_i(n)$, independently from the
choices of the other balls.
Moreover suppose that for all $i$ ($0\le i\le k$) we have $p_i=o(1)$,
where asymptotics are with respect to $n\to\infty$ and $k$ is assumed to
be fixed.
Then, the probability $P$ that $\cU_0$ contains no balls but each
$\cU_i$ has at least one for all $1\le i\le k$ is
\[
P\sim (1-p_0)^n \prod_{i=1}^k (1-e^{-np_i}).
\]
\end{lem}
\begin{proof}
By reordering the labels of the bins except for $\cU_0$, assume that
$np_i=o(1)$ if $1\le i\le r$ and $np_i=\Omega(1)$ if $r+1\le i\le k$.
For any $f=o(1)$, we define
\[
P_f = \sum_{a_1,\ldots,a_r} (-1)^{\sum_{i=1}^ra_i}
\left(1-\frac{\sum_{i=1}^ra_ip_i}{1-f}\right)^n,
\]
where the summation indices $a_1,\ldots,a_r$ run from $0$ to $1$. We can
write
\[
P_f = \sum_{a_1,\ldots,a_r} (-1)^{\sum_{i=1}^ra_i}
\sum_{\substack{m_0,\ldots,m_r\ge0\\m_0+\cdots+m_r= n}}
\binom{n}{m_0,\ldots,m_r} \prod_{i=1}^r
\left(\frac{-a_ip_i}{1-f}\right)^{m_i},
\]
with the convention $0^0=1$. A changing of the order of
summation converts this expression to
\[
P_f = \sum_{\substack{m_0,\ldots,m_r\ge0\\m_0+\cdots+m_r= n}}
\binom{n}{m_0,\ldots,m_r} \prod_{i=1}^r
\left(\frac{-p_i}{1-f}\right)^{m_i}
\sum_{a_1,\ldots,a_r} (-1)^{\sum_{i=1}^ra_i}
\prod_{i=1}^r a_i^{m_i}.
\]
All terms in this sum with some $m_i=0$ ($1\le i\le r$) cancel, since
each of these terms is equal but has opposite sign to the one obtained
by switching the value of $a_i$. So, only the terms with all
$m_1,\ldots,m_r\ge 1$ remain, and among these we can remove the ones
with some $a_i=0$. Hence,
\[
P_f = \sum_{\substack{m_0\ge0\\m_1\ldots,m_r\ge1\\m_0+\cdots+m_r= n}}
(-1)^r \binom{n}{m_0,\ldots,m_r} \prod_{i=1}^r
\left(\frac{-p_i}{1-f}\right)^{m_i}.
\]
Since $np_i=o(1)$, the main asymptotic weight in this sum corresponds to
the term $m_0=n-r$ and $m_1,\ldots,m_r=1$, so
\begin{equation}\label{eq:Pf}
P_f \sim \frac{[n]_r}{(1-f)^r} \prod_{i=1}^r p_i \sim \prod_{i=1}^r np_i
\sim \prod_{i=1}^r (1-e^{-np_i}).
\end{equation}
By an inclusion-exclusion argument, the probability in the statement can
be written as
\[
P = \sum_{a_1,\ldots,a_k} (-1)^{\sum_{i=1}^k a_i}
\left(1-p_0-\sum_{i=1}^k a_ip_i \right)^n,
\]
where the summation indices $a_1,\ldots,a_k$ run from $0$ to $1$. Then,
if we define
\[
P_{a_{r+1},\ldots,a_k} = \sum_{a_1,\ldots,a_r} (-1)^{\sum_{i=1}^r a_i}
\left(1-\frac{\sum_{i=1}^r a_ip_i}{1-p_0-\sum_{i=r+1}^k a_ip_i} \right)^n,
\]
we can write
\begin{align}
P &= (1-p_0)^n \sum_{a_{r+1},\ldots,a_k} (-1)^{\sum_{i=r+1}^k a_i}
\left(1-\frac{\sum_{i=r+1}^k a_ip_i}{1-p_0} \right)^n P_{a_{r+1},\ldots,a_k}
\notag\\
&= (1-p_0)^n \sum_{a_{r+1},\ldots,a_k} (-1)^{\sum_{i=r+1}^k a_i}
\exp\left(-(1+o(1))\sum_{i=r+1}^k a_inp_i \right) P_{a_{r+1},\ldots,a_k}.
\label{eq:theP}
\end{align}
Note that for each $a_{r+1},\ldots,a_k\in\{0,1\}$, in view
of~\eqref{eq:Pf} and setting $f=p_0+\sum_{i=r+1}^k a_ip_i$, we have
\begin{equation}\label{eq:Paa}
P_{a_{r+1},\ldots,a_k} \sim \prod_{i=1}^r (1-e^{-np_i}).
\end{equation}
The fact that $np_i=\Omega(1)$ for $r+1\le i\le m$ prevents the leading
term of the sum in~\eqref{eq:theP} from cancelling out. Thus,
from~\eqref{eq:theP} and~\eqref{eq:Paa}, we obtain
\[
P \sim (1-p_0)^n \prod_{i=1}^k (1-e^{-np_i}).
\qedhere
\]
\end{proof}

We are now in good position to study the changes experienced by the isolated vertices between two consecutive steps $t$ and $t+1$. Extending the notation in Section~\ref{sec:static}, we denote by $K_{1,t}$ the number of isolated vertices of $\RGt$.
Also, for any two consecutive steps $t$ and $t+1$, we define the following random variables:
$B_t$ is the number of agents $i$ such that $X_i$ is not isolated in $\RGt$ but $X'_i$ is isolated in $\RGtt$;
$D_t$ is the number of agents $i$ such that $X_i$ is isolated in $\RGt$ but $X'_i$ is not isolated in $\RGtt$;
$S_t$ is the number of agents $i$ such that $X_i$ and $X'_i$ are both isolated in $\RGt$ and $\RGtt$.
For simplicity, we often denote them just by $B$, $D$ and $S$ whenever $t$ and $t+1$ are understood. Note that $B$ and $D$ have the same distribution, since any creation of an isolated vertex corresponds to a destruction of an isolated vertex in the time-reversed process and vice versa.

To state the following result, we need one more definition:
Given a collection of events $\cE_1(n),\ldots,\cE_k(n)$ and of random variables $W_1(n),\ldots,W_l(n)$ taking values in $\nat$, with $k$ and $l$ fixed, we say that they are mutually asymptotically independent if for any $k',l',i_1,\ldots,i_{k'},j_1,\ldots,j_{l'},w_1,\ldots,w_{l'}\in\nat$ such that $k'\le k$, $l'\le l$, $1\le i_1<\cdots<i_{k'}\le k$, $1\le j_1<\cdots<j_{l'}\le l$ we have that
\begin{equation} \label{eq:asymp_ind}
\pr\left( \bigwedge_{a=1}^{k'}\cE_{i_a} \wedge \bigwedge_{b=1}^{l'} (W_{j_b}=w_b) \right)
\sim \prod_{a=1}^{k'}\pr(\cE_{i_a}) \prod_{b=1}^{l'}\pr(W_{j_b}=w_b).
\end{equation}
%
%
\begin{prop}\label{prop:BDS}
Assume $\mu=\Theta(1)$. Then for any two consecutive steps,
\[
\ex B = \ex D \sim \mu (1-e^{-qn}) \quad\text{and}\quad \ex S \sim \mu e^{-qn}.
\]
Moreover we have that
\begin{enumerate}
\item If $s=o(1/rn)$, then $\pr(B>0)\sim\ex B$; $\pr(D>0)\sim\ex D$; $S$ is asymptotically Poisson; and $(B>0)$, $(D>0)$ and $S$ are asymptotically mutually independent.
\item If $s=\Theta(1/rn)$, then $B$, $D$ and $S$ are asymptotically mutually independent Poisson.
\item If $s=\omega(1/rn)$, then $B$ and $D$ are asymptotically Poisson; $\pr(S>0)\sim\ex S$; and $B$, $D$ and $(S>0)$ are asymptotically mutually independent.
\end{enumerate}
\end{prop}
\begin{proof}
The central ingredient in the proof is the computation of the joint factorial moments
$\EXP{[B]_{\ell_1}[D]_{\ell_2}[S]_{\ell_3}}$ of these variables. In particular, 
we find the asymptotic values of $\EXP{B}$, $\EXP{D}$ and $\EXP{S}$. Moreover, in the case $s=\Theta\big(1/(rn)\big)$, we show that for any fixed naturals $\ell_1$, $\ell_2$ and $\ell_3$ we have
\begin{equation} \label{eq:moments2}
\EXP{[B]_{\ell_1}[D]_{\ell_2}[S]_{\ell_3}} \sim \EXP{B}^{\ell_1}
\EXP{D}^{\ell_2} \EXP{S}^{\ell_3}.
\end{equation}
Then, the result follows from Theorem 1.23 in~\cite{Bollobas01}.
The other cases are more delicate since~\eqref{eq:moments2} does not always hold for extreme values of $s$, and we obtain a weaker result. In the case $s=o\big(1/(rn)\big)$, we compute the moments for any natural $\ell_3$ but only for $\ell_1,\ell_2\in\{0,1,2\}$ and obtain
\begin{align}
&\EXP{[B]_{\ell_1}[D]_{\ell_2}[S]_{\ell_3}} \sim (\EXP{B}^{\ell_1}
(\EXP{D})^{\ell_2} (\EXP{S})^{\ell_3}, \quad \text{if }\ell_1,\ell_2<2,
\notag\\
&\EXP{[B]_{\ell_1}[D]_{\ell_2}[S]_{\ell_3}}  = o(\EXP{B\,[D]_{\ell_2}[S]_{\ell_3}}),
\notag\\
&\EXP{[B]_{\ell_1}[D]_{\ell_2}[S]_{\ell_3}}  = o(\EXP{[B]_{\ell_1}D\,[S]_{\ell_3}}).
    \label{eq:moments3}
\end{align}
{From} this and by using upper and lower bounds given in~\cite{Bollobas01},
Section 1.4, applied to several variables, we deduce that $(B>0)$, $(D>0)$ and $S$ satisfy~\eqref{eq:asymp_ind} and also
\[
\PR{B>0} \sim \EXP{B}, \quad
\PR{D>0} \sim \PR{D} \quad\text{and}\quad
\PR{S=k} \sim e^{-\EXP{S}}\frac{\EXP{S}^k}{k!} \quad \forall k\in\nat.
\]
Similarly, in the case $s=\omega\big(1/(rn)\big)$, we compute the moments for any naturals $\ell_1$ and $\ell_2$ but only for $\ell_3\in\{0,1,2\}$ and obtain
\begin{align}
&\EXP{[B]_{\ell_1}[D]_{\ell_2}[S]_{\ell_3}}   \sim (\EXP{B})^{\ell_1}
(\EXP{D})^{\ell_2} (\EXP{S})^{\ell_3}, \quad \text{if }\ell_3<2,
\notag\\
&\EXP{[B]_{\ell_1}[D]_{\ell_2}[S]_{\ell_3}}   = o(\EXP{[B]_{\ell_1}[D]_{\ell_2}S})
    \label{eq:moments4}
\end{align}
{From} this and by using once more upper and lower bounds given in Section 1.4 of~\cite{Bollobas01}, we conclude that $B$, $D$ and $(S>0)$ satisfy~\eqref{eq:asymp_ind} and also
\begin{gather*}
\PR{B=k} \sim e^{-\EXP{B}}\frac{(\EXP{B})^k}{k!} \quad \forall k\in\nat,
\\
\PR{D=k} \sim e^{-\EXP{D}}\frac{(\EXP{D})^k}{k!} \quad \forall k\in\nat \quad\text{and}\quad
\PR{S>0} \sim \EXP{S}.
\end{gather*}
First, we define for each $i\in\nset$ $B_i$, $D_i$ and $S_i$ as the indicator functions of the following events respectively: $X_i$ is not isolated in $\RGt$ but $X'_i$ is isolated in $\RGtt$; $X_i$ is isolated in $\RGt$ but $X'_i$ is not isolated in $\RGtt$; $X_i$ and $X'_i$ are both isolated in $\RGt$ and $\RGtt$. This allows us to write
\[
B=\sum_{i=1}^n B_i, \quad D=\sum_{i=1}^n D_i, \quad S=\sum_{i=1}^n S_i.
\]
Note that $B_i=1$ iff all points in $\hcX\setminus\{\hX_i\}$ are outside $\hcR'_i$ but at least one is inside $\hcQ'_i$; also $D_i=1$ iff all points in $\hcX\setminus\{\hX_i\}$ are outside $\hcR_i$ but at least one is inside $\hcQ_i$; and finally $S_i=1$ iff all points in $\hcX\setminus\{\hX_i\}$ are outside $\hcR_i\cup\hcR'_i=\hcR_i\cup\hcQ_i=\hcR'_i\cup\hcQ'_i$.

Now given any fixed naturals $\ell_1,\ell_2,\ell_3$ with sum $\ell=\ell_1+\ell_2+\ell_3$,
we choose an ordered tuple $J$ of $\ell$ different vertices $i_1,\ldots,i_\ell\in\nset$,
and define
\begin{equation}\label{eq:Ej}
\cE = \bigwedge_{a=1}^{\ell_1} (B_{i_a}=1) \wedge \bigwedge_{b=\ell_1+1}^{\ell_1+\ell_2} (D_{i_b}=1) \wedge \bigwedge_{c=\ell_1+\ell_2+1}^{\ell} (S_{i_c}=1).
\end{equation}
\remove{
namely the event that during two consecutive time steps $t$ and $t+1$ and for some particular $\ell$ vertices, the first $\ell_1$ become isolated, the next $\ell_2$ stop being isolated and the last $\ell_3$ stay isolated.
}
Observe that $\pr(\cE)$ does not depend on the particular tuple $J$, and multiplying it by $[n]_\ell$ (i.e.\ the number of ordered choices of $J$) we get
\begin{equation}
	\label{eq:moments}
\ex([B]_{\ell_1}[D]_{\ell_2}[S]_{\ell_3}) = [n]_\ell \pr(\cE)
\end{equation}
By relabelling the vertices in $J$ we assume hereinafter that $J=(1,\ldots,\ell)$, and we call $\hcY=\bigcup_{i=1}^\ell\{\hX_{i}\}$.
Moreover, we define the set
\[
\hcR =\bigcup_{i=1}^{\ell_1}\hcR'_{i} \cup \bigcup_{i=\ell_1+1}^{\ell_1+\ell_2}\hcR_{i} \cup \bigcup_{i=\ell_1+\ell_2+1}^{\ell}(\hcR_{i}\cup\hcR'_{i})
\]
and the collection of sets
\[
\bfm{\hcQ} = \{\hcQ'_1,\ldots,\hcQ'_{\ell_1},\hcQ_{\ell_1+1},\ldots,\hcQ_{\ell_1+\ell_2} \},
\]
which play an important role in the computation of $\pr(\cE)$.
It is useful to call $\hcQ^*_{i}=\hcQ'_{i}$ for $1\le i\le \ell_1$,
$\hcQ^*_{i}=\hcQ_{i}$ for $\ell_1+1\le i\le \ell_1+\ell_2$, so that we can write $\bfm{\hcQ} = \{\hcQ^*_1,\ldots,\hcQ^*_{\ell_1+\ell_2}\}$.
\setcounter{case}{0}
\begin{case}[$\,s=\Theta\big(1/(rn)\big)\,$]
We say that a vertex $i\in J$ is {\em restricted} if there is some other $j\in J$ with $j>i$ such that $d(X_{i},X_{j})\le 2r+4s$.
Let $\cF$ be the event that there are no restricted vertices in $J$, i.e.\ $d(X_{i},X_{j})>2r+4s$ for all $i,j\in J$ ($i\ne j$). This has probability $1-O(r^2)$. We first suppose that  $\cF$ holds and compute the probability of $\cE$ conditional upon that.
We observe that $\cF$ implies that for any $i,j\in J$ ($i\ne j$) we must have
$\hcR_{i} \cap \hcR_{j}=\emptyset$, $\hcR'_{i} \cap \hcR'_{j}=\emptyset$ and $\hcR_{i} \cap \hcR'_{j}=\emptyset$.
Then $\vol{\hcR}=\ell\pi r^2+\ell_3q$, and the sets in $\bfm\hcQ$ are pairwise disjoint and also disjoint from $\hcR$.
Moreover observe that, conditional upon $\cF$, $\cE$ is equivalent to the event that all points in $\hcX\setminus\hcY$ lie outside $\hcR$, but at least one belongs to each $\hcQ^*_i\in\bfm\hcQ$. From all the above, the probability of $\cE$ can be easily obtained by a Balls and Bins argument. In fact, from Lemmata~\ref{lem:q} and~\ref{lem:inclusion_exclusion} we conclude
\begin{align}
\pr(\cE\wedge\cF) &= (1-O(r^2)) \, \pr(\cE\mid\cF)
\notag\\
& \sim (1-\ell\pi r^2-\ell_3q)^n(1-e^{-qn})^{\ell_1+\ell_2}
\notag\\
&\sim \left(\frac\mu n\right)^\ell (1-e^{-qn})^{\ell_1+\ell_2} e^{-\ell_3qn}.
	\label{eq:prEjF}
\end{align}
We claim that this is the main contribution to $\pr(\cE)$. In fact if $\cF$ does not hold (i.e.\ some of the points in $\hcY$ are at distance at most $2r+4s$), then $\pr(\cE\mid\overline\cF)$ is larger than the expression in~\eqref{eq:prEjF}, but this is balanced out by the fact that $\pr(\overline\cF)$ is small.
Before proving this claim, define $\cH$ to be the event that  $d(X_{i},X_{j})>r-2s$ for all $i,j\in J$ ($i\ne j$). Notice that $\cE$ implies $\cH$, since otherwise, for some $i,j\in J$, $X_{i}$ and $X_{j}$ would be joined by an edge in $\RGt$ and also $X'_{i}$ and $X'_{j}$ in $\RGtt$, which is not compatible with $\cE$. Therefore we only need to see that $\pr(\cE\wedge\overline\cF) = \pr(\overline\cF\wedge\cH) \pr(\cE\mid\overline\cF\wedge\cH)$ is negligible compared to~\eqref{eq:prEjF}.

Suppose then that $\cH$ holds and also that $p>0$ of the vertices in $J$ are restricted (i.e.\ $\cF$ does not hold). This happens with probability $O(r^{2p})$.
In this case, we deduce that $\vol{\hcR}\ge (\ell-p)\pi r^2 + \epsilon\pi r^2$, since each unrestricted vertex in $J$ contributes at least $\pi r^2$ to $\vol{\hcR}$ and the first restricted one gives by Lemma~\ref{lem:tec}(2) the term $\epsilon\pi r^2$.
Moreover,  $\cE$ implies that all points in $\hcX\setminus\hcY$ lie outside of $\hcR$, which has probability $\big(1-\vol{\hcR}\big)^{n-\ell}=O(1/n^{\ell-p+\epsilon})$.
Summarizing, the weight in $\pr(\cE\wedge\overline\cF)$ coming from situations with $p$ restricted vertices is $O(r^{2p}/n^{\ell-p+\epsilon})=O(\log^p n/n^{\ell+\epsilon})$, and is thus negligible compared to~\eqref{eq:prEjF}. Hence $\pr(\cE)\sim\pr(\cE\wedge\cF)$, and the required condition on the moments announced in~\eqref{eq:moments2} follows from~\eqref{eq:moments} and~\eqref{eq:prEjF}.
\end{case}
\begin{case}[$\,s=o\big(1/(rn)\big)\,$]
Defining $\cF$ and $\cH$ as in the case $s=\Theta\big(1/(rn)\big)$
and by an analogous argument we obtain
\begin{equation}
	\label{eq:prEjF2}
\pr(\cE\wedge\cF) \sim \left(\frac\mu n\right)^\ell (1-e^{-qn})^{\ell_1+\ell_2} e^{-\ell_3qn} \sim \left(\frac\mu n\right)^\ell (qn)^{\ell_1+\ell_2}
\end{equation}
However, the analysis of the case that $\cF$ does not hold is slightly more delicate here.
Indeed, there is an additional $o(1)$ factor in~\eqref{eq:prEjF2}, namely $(qn)^{\ell_1+\ell_2}$, which forces us to get tighter bounds on $\pr(\cE\wedge\overline\cF\wedge\cH)$ than the ones obtained before.
Unlike in the case $s=\Theta\big(1/(rn)\big)$, here we need to consider also the role of 
$\bfm\hcQ$ when $\cF$ does not hold, and special care must be taken with several new situations which do not 
occur otherwise. For instance, since the elements of $\bfm\hcQ$ are not necessarily disjoint, then 
for $\hcQ^*_{i},\hcQ^*_{j}\in\bfm\hcQ$ the condition that both contain some element of $\hcX$ can 
be satisfied by having just a single point in $\hcQ^*_{i}\cap\hcQ^*_{j}\cap\hcX$. Moreover, 
if $\ell_1\ge2$ and $1\le i < j\le \ell_1$ (or $\ell_2\ge2$ and $\ell_1+1\le i,j\le \ell_1+\ell_2$), the previous condition is also satisfied if $\hX_{j}\in\hcQ^*_{i}$, which is equivalent to $\hX_{i}\in\hcQ^*_{j}$. If the latter situation occurs, we say that $i$ and $j$ \emph{collaborate}.

We now distinguish two cases whether two vertices both belonging to $J$ collaborate 
or a vertex from outside $J$ causes $\overline\cF$ not to hold. First, we 
bound the weight in $\pr(\cE\wedge\overline\cF)$ due to situations in which there are no pairs of elements 
in $J$ which collaborate.
We need some definitions. Let $J_1=\{1,\ldots,\ell_1+\ell_2\}$ and $\hcY_1=\bigcup_{i=1}^{\ell_1+\ell_2} \{\hX_i\}$, and consider the class $\mathscr P$ of partitions of $J_1$.
Namely, a partition of $J_1$ is a collection of \emph{blocks} (nonempty subsets of $J_1$) which are disjoint and have union $J_1$. The size of a partition is the number of blocks, and for each block we call \emph{leader} to the maximal element in the block.
Given a partition $P=\{A_1,\ldots,A_k\}\in\mathscr P$ and also $i_1,\ldots,i_k\in\nset\setminus J$, let $\EPi$ be the following event: For each block $A_j$ of $P$, we have $\hX_{i_j}\in\bigcap_{i\in A_j} \hcQ^*_{i}$ and moreover all the points in $\hcX\setminus(\hcY\cup\{i_1,\ldots,i_k\})$ lie outside of $\hcR$. We wish to bound the probability of $\EPi\wedge\overline\cF\wedge\cH$.
Notice that if $\EPi$ holds, then all the $\ell_1+\ell_2-k$ non-leader elements in $J_1$ must be restricted, and possibly some other $p'$ vertices in $J$ are restricted too.
Moreover, $\cF$ does not hold iff this $p'$ satisfies $0<\ell_1+\ell_2-k+p'<\ell$.
Given any $p'$ with that property, suppose that the number of restricted vertices in $J$ which are either in $J\setminus J_1$ or are leaders of some block is exactly $p'$. We condition upon this and also upon $\cH$, which has probability $r^{2p'}$.
Then for each block $A_j$ with leader $l_j$, event $\EPi$ requires that $\hX_{i_j}\in\hcQ^*_{l_j}$ and for all $i\in A_j$ ($i\neq l_j$) $\hX_{i}\in(\hcQ_{i_j}\cup\hcQ'_{i_j})$.
In addition, since the number of restricted vertices in $J$ is $\ell_1+\ell_2-k+p'>0$, arguing as in the case $s=\Theta\big(1/(rn)\big)$ we have $\vol{\hcR}\ge (\ell_3+k-p')\pi r^2 + \epsilon\pi r^2$. Then the contribution to $\pr(\EPi\wedge\overline\cF\wedge\cH)$ for this particular $p'$ is
\[
O(r^{2p'}) q^k (2q)^{\ell_1+\ell_2-k} (1-\vol{\hcR})^{n-\ell-k}
= O\left(\frac{\log^{p'}n}{n^{\ell+k+\epsilon}}\right) (qn)^{\ell_1+\ell_2},
\]
so for some $0<\epsilon'<\epsilon$ we can write
\[
\pr(\EPi\wedge\overline\cF\wedge\cH) = O\left(\frac{1}{n^{\ell+k+\epsilon'}}\right) (qn)^{\ell_1+\ell_2}.
\]
Finally observe that if there are no pairs of elements in $J$ which collaborate, then $\cE\wedge\overline\cF$ implies that $\EPi\wedge\overline\cF\wedge\cH$ holds for some $P\in\mathscr P$ of size $k$ and some $i_1,\ldots,i_k\in\nset\setminus J$, and therefore has  probability
\begin{equation}\label{eq:PEnF1}
O\left(n^k\right) O\left(\frac{1}{n^{\ell+k+\epsilon'}}\right) (qn)^{\ell_1+\ell_2}
= O\left(\frac{1}{n^{\ell+\epsilon'}}\right) (qn)^{\ell_1+\ell_2},
\end{equation}
negligible compared to~\eqref{eq:prEjF2}.
In particular, if $\ell_1,\ell_2<2$, then no pair of elements in $J$ collaborates and then $\pr(\cE)\sim\pr(\cE\wedge\cF)$. Hence, the first line of~\eqref{eq:moments3} follows from~\eqref{eq:moments} and~\eqref{eq:prEjF2}.

We now extend the approach above to deal with situations in which some pair of elements in $J$ collaborate. Unfortunately, their contribution to $\pr(\cE\wedge\overline\cF\wedge\cH)$ may be larger than~\eqref{eq:prEjF2} if $s$ tends to $0$ fast. Hence we restrict $\ell_1$ and $\ell_2$ to be at most $2$ and prove only~\eqref{eq:moments3}.
If $\ell_1=2$ let $\cE_1$ be the following event: $\hX_1\in\hcQ'_2$, $\hcR$ contains no points in $\hcX\setminus\hcY$ and also, for each natural ($3\le i\le 2+\ell_2$), $\hcQ_i$ contains some point in $\hcX\setminus\hcY$.
Similarly if $\ell_2=2$ let $\cE_2$ be the following event: $\hX_{\ell_1+1}\in\hcQ_{\ell_1+2}$, $\hcR$ contains no points in $\hcX\setminus\hcY$ and also, for each natural $i$ ($1\le i\le\ell_1$), $\hcQ'_i$ contains some point in $\hcX\setminus\hcY$.
Finally if $\ell_1=\ell_2=2$ let $\cE_{1,2}$ be the following event: $\hX_{1}\in\hcQ'_{2}$ and also $\hX_{3}\in\hcQ_{4}$.
In order to compute $\pr(\cE_1\wedge\cH)$, we can repeat the same argument above, but imposing that $\hX_1\in\hcQ'_2$ and ignoring other conditions on $\hcQ'_1$ and $\hcQ'_2$. We obtain for some $\epsilon'>0$
\begin{equation}\label{eq:PE1H}
\pr(\cE_1\wedge\cH) = O\left(\frac{1}{n^{\ell-1+\epsilon'}}\right) q (qn)^{\ell_2}
= O\left(\frac{1}{n^{\ell+\epsilon'}}\right) (qn)^{1+\ell_2},
\end{equation}
and similarly
\begin{equation}\label{eq:PE2H}
\pr(\cE_2\wedge\cH) = O\left(\frac{1}{n^{\ell+\epsilon'}}\right) (qn)^{\ell_1+1}
\quad\text{and}\quad
\pr(\cE_{1,2}\wedge\cH)  = O\left(\frac{1}{n^{\ell+\epsilon'}}\right) (qn)^2.
\end{equation}
Observe that if some vertices in $J$ collaborate, then $\cE\wedge\overline\cF$ implies
that $\cE_1\wedge\cH$, $\cE_2\wedge\cH$ or $\cE_{1,2}\wedge\cH$ hold. Unfortunately, from~\eqref{eq:prEjF2}, \eqref{eq:PE1H} and~\eqref{eq:PE2H} we cannot guarantee that $\pr(\cE\wedge\overline\cF)$ is smaller than $\pr(\cE\wedge\cF)$, but in any case, by multiplying these probabilities by $[n]_\ell$ in view of~\eqref{eq:moments}, we complete the proof of~\eqref{eq:moments3}.
\end{case}
\begin{case}[$\,s=\omega\big(1/(rn)\big)$ but also $s=O(r)\,$]
Following the same notation as in the case $s=\Theta\big(1/(rn)\big)$ and by an analogous argument we obtain
\begin{equation}
	\label{eq:prEjF3}
\pr(\cE\wedge\cF) \sim \left(\frac\mu n\right)^\ell (1-e^{-qn})^{\ell_1+\ell_2} e^{-\ell_3qn} \sim \left(\frac\mu n\right)^\ell e^{-\ell_3qn}
\end{equation}
If $\ell_3\le1$, we claim that this is the main contribution to $\pr(\cE)$. In fact, suppose that $\cH$ holds and also that $p>0$ of the vertices in $J$ are restricted (i.e.\ $\cF$ does not hold). This happens with probability $O(r^{2p})$. Since $\ell_3\le1$, then the only possible event which contributes to $S$ required in the definition of $\cE$ is $(S_{\ell})$, involving vertex $\ell$ which cannot be restricted by definition.
Then we deduce that $\vol{\hcR}\ge (\ell-p)\pi r^2 + \ell_3q + \epsilon\pi r^2$, since the unrestricted vertices in $J$ contribute $(\ell-p)\pi r^2+\ell_3q$ to $\vol{\hcR}$ and the first restricted one gives the term $\epsilon\pi r^2$, by Lemma~\ref{lem:tec}(2--3). Therefore, the probability of $\cE$ in this situation is $O(e^{-\ell_3qn}/n^{\ell-p+\epsilon})$, which combined with the probability $O(r^{2p})$ that $p$ vertices are restricted has negligible weight compared to~\eqref{eq:prEjF3}. Hence, $\pr(\cE)\sim\pr(\cE\wedge\cF)$, and the first line of~\eqref{eq:moments4} follows from~\eqref{eq:moments} and~\eqref{eq:prEjF3}.

Unfortunately, if $\ell_3=2$ and we have $p$ restricted vertices in $J$, we can only assure that
$\vol{\hcR}\ge (\ell-p)\pi r^2 + q + \epsilon\pi r^2$, and then for some $0<\epsilon'<\epsilon$
\begin{equation}	\label{eq:PEnF2}
\pr(\cE\wedge\overline\cF) = O\left(\frac{r^{2p}}{n^{\ell-p+\epsilon}}\right) e^{-qn}
= O\left(\frac{1}{n^{\ell+\epsilon'}}\right) e^{-qn},
\end{equation}
which may have significant contribution to $\pr(\cE)$ if $s$ is large enough. But in any case, in view of~\eqref{eq:moments}, \eqref{eq:prEjF3} and~\eqref{eq:PEnF2}, we verify that the second line of~\eqref{eq:moments4} is satisfied.
\end{case}
\begin{case}[$\,s=\omega(r)\,$]
Let $\cF'$ be the event that for any $i,j\in J$ ($i\ne j$) we have $d(X_i,X_j)>2r$ and also $d(X'_i,X'_j)>2r$. This event has probability $1-O(r^2)$. We observe that if $\cF'$ holds,
then for any $i,j\in J$ ($i\ne j$) we must have $\hcR_i\cap\hcR_j=\emptyset$, $\hcR'_i\cap\hcR'_j=\emptyset$ and $\hcR_i\cap\hcR'_j=\emptyset$. Therefore,
$\vol{\hcR}=\ell\pi r^2+\ell_3q$ and the sets in $\bfm\hcQ$ are pairwise disjoint and also disjoint from $\hcR$. Then in view of Lemmata~\ref{lem:q} and~\ref{lem:inclusion_exclusion}, and by the same argument that leads to~\eqref{eq:prEjF}
\begin{equation}
	\label{eq:prEjF4}
\pr(\cE\wedge\cF') \sim \left(\frac\mu n\right)^\ell (1-e^{-qn})^{\ell_1+\ell_2} e^{-\ell_3qn} \sim \left(\frac\mu n\right)^\ell e^{-\ell_3qn}
\end{equation}
The remaining of the argument is analogous to the previous case but replacing $\cF$ with $\cF'$ and using Lemma~\ref{lem:tec}(4).
\qedhere
\end{case}
\end{proof}
Taking into account that $K_{1,t}=D_t+S_t$ and $K_{1,t+1}=S_{t}+B_{t}$,  
the number of isolated vertices at two consecutive steps can in the case $s=\Theta\big(1/(rn)\big)$ 
be completely characterized by Proposition~\ref{prop:BDS}. For the other ranges of $s$, 
the result is weaker but still sufficient for our further purposes. We remark that if $s=o\big(1/(rn)\big)$ 
then creations and destructions of isolated vertices are rare, but a Poisson number 
of isolated vertices is present at both consecutive steps. Otherwise if $s=\omega\big(1/(rn)\big)$ 
then the isolated vertices which are present at both consecutive steps are rare since, but a Poisson number of them 
is created and also a Poisson number destroyed.

Now in order to characterize the connectivity of \RGT, we need to bound the probability that components other than isolated vertices and the giant one appear at some step. We know by Theorem~\ref{thm:static} that a.a.s.\ this does not occur at one single step $t$. However during long periods of time this event could affect the connectivity and must be considered.

Extending the notation in Section~\ref{sec:static}, given a step $t$ let $\tK_{2,t}$ be the number of non-solitary components other than isolated vertices occurring at step $t$. We show that they have a negligible effect compared to isolated vertices in the dynamic evolution of connectivity.
%
%
\begin{lem}\label{lem:no_birth_medium}
Assume that $\mu=\Theta(1)$ and $s=o\big(1/(rn)\big)$. Then,
\begin{itemize}
\item $\pr(\tK_{2,t} >0 \wedge \tK_{2,t+1}=0) = \pr(\tK_{2,t} =0 \wedge \tK_{2,t+1}>0) =o(srn),$
\item $\pr(\tK_{2,t} >0 \wedge B_t>0)=o(srn).$
\end{itemize}
\end{lem}
\begin{proof}
Recall from Lemma~\ref{lem:q} that  if $s=o\big(1/(rn)\big)$ then $q=\Theta(rs)$.
Then it is enough to prove that $\pr(\tK_{2,t} >0 \wedge \tK_{2,t+1}=0) = o(qn)$ and $\pr(\tK_{2,t} >0 \wedge B_t>0)=o(qn)$, since $(\tK_{2,t} =0 \wedge \tK_{2,t+1}>0)$ corresponds in the time-reversed process to $(\tK_{2,t} >0 \wedge \tK_{2,t+1}=0)$ and thus they have the same probability.

Consider all the possible components in $\nRG$ which are not solitary and have size at least $2$. They are classified into several types according to their size and diameter, and we deal with each type separately. Then if we denote by $M_i$ the number of components of type~$i$ in \RGt, we must show for each $i$ that
\begin{equation}\label{eq:PMi}
\pr(M_i >0 \wedge \tK_{2,t+1}=0) = o(qn) \quad\text{and}\quad \pr(M_i >0 \wedge B_t>0)=o(qn).
\end{equation}

Also we need one definition which helps to describe the changes of edges between \RGt\ and \RGtt. For each $i\in\nset$ we define $\hcP_i=\hcQ_i\cup\hcQ'_i=\hcR_i\Delta\hcR'_i$ (where $\Delta$ denotes the symmetric difference of sets). Given also $j\in\nset$, see that $\hX_j\in\hcP_i$ iff $\hX_i\in\hcP_j$ iff vertices $i$ and $j$ share an edge either at time $t$ or at time $t+1$ but not at both times (which happens with probability $\vol{\hcP_i}=2q$).

Each part in this proof is labelled by a number followed by a prime ($'$) in order to avoid confusion with the parts in the proof of Lemma~\ref{lem:PY}, which are often referred to. Moreover, we write for simplicity Part~$i$ (p.L.~\ref{lem:PY}) to denote Part~$i$ in the proof of Lemma~\ref{lem:PY}.

We set throughout this proof  $\epsilon=10^{-18}$.
\setcounter{pt}{0}
\renewcommand{\thept}{\arabic{pt}$'$}
\begin{pt}\label{p:small2}
Consider all the possible components in $\nRG$ which have diameter at most $\epsilon r$ and 
size between $2$ and  $\log n/37$. Call them components of type 1, and let $M_1$ 
denote their number at time $t$. This definition is similar to the one in Part~1 
(p.L.~\ref{lem:PY}), but also includes components of size $2$, covered by Lemma~\ref{lem:EZei}.

\remove{
First observe that, given a fixed $p\in\nat$ and $i_1,\ldots,i_p\in\nset$, the arguments in the proofs of Lemmata~\ref{lem:EZei} and~\ref{lem:PY} are still valid if we replace $\cX$ by $\cX\setminus\{X_{i_1},\ldots,X_{i_p}\}$ (i.e.\  we ignore a fixed set  of vertices in the model).
Hence, the probability of having some component in $G(\cX\setminus\{X_{i_1},\ldots,X_{i_p}\};r)$ of type~\ref{p:small} and size at least $\ell\ge2$ is $O(1/\log^{\ell-1}n)$. We will use this extension several times in this argument.
}

Given any $i\in\nset$, let $\cE_i$ be the following event: 
There exists a component $\Gamma$ of type~1 in $G(\cX\setminus\{X_i\};r)$ and moreover for some $j\in\nset$ such that $X_j$ is a vertex of $\Gamma$ we have that $\hX_i\in\hcP_j$.
In order to compute the probability of $\cE_i$, we note that
the arguments in the proofs of Lemmata~\ref{lem:EZei} and~\ref{lem:PY} are still valid if we replace $\cX$ by $\cX\setminus\{X_{i}\}$ (i.e.\  we ignore vertex $i$ in the model).
Hence, the probability of having some component in $G(\cX\setminus\{X_{i}\};r)$ of type~1 
and size at least $\ell\ge2$ is $O(1/\log^{\ell-1}n)$.
Suppose first that $G(\cX\setminus\{X_i\};r)$ has some component $\Gamma$ of type~1 and size between $3$ and $\log n/37$. This happens with probability $O(1/\log^2 n)$. Conditional upon this, the probability that $\hX_i\in\hcP_j$ for some $j\in\nset$ with $X_j$ being a vertex of $\Gamma$ is at most $\log n/37$ times $2q$. This 
contributes $O(1/\log^2 n) (\log n/37)(2q) = O(q/\log n)$ to the probability of $\cE_i$.
Otherwise suppose that $G(\cX\setminus\{X_i\};r)$ has some component $\Gamma$ of type~1 and size exactly $2$. This happens with probability $O(1/\log n)$.
Conditional upon this, the probability that $\hX_i\in\hcP_j$
for some $j\in\nset$ with $X_j$ being a vertex of $\Gamma$ is at most two times $2q$. This also contributes 
$O(1/\log n) (4q) = O(q/\log n)$ to the probability of $\cE_i$, and therefore $\pr(\cE_i)=O(q/\log n)$.

Given any $i_1,i_2\in\nset$ ($i_1\ne i_2$), let $\cF_{i_1,i_2}$ be the following event:
There exists a component $\Gamma$ of type~1 in $G(\cX\setminus\{X_{i_2}\};r)$ and moreover $\hcR'_{i_1}\cap(\hcX\setminus\{\hX_{i_1},\hX_{i_2}\})=\emptyset$.
To derive the probability of $\cF_{i_1,i_2}$, we distinguish two cases according to the distance between $X_{i_1}$ and $\Gamma$.
Suppose first that for some $h\in\nset\setminus\{i_1,i_2\}$ we have that $r<d(X_{i_1},X_h)\le 3r$ (which happens with probability $O(r^2)=O(\log n/n)$). Let $\cS_h$ be the set of points in the torus \UT\ at distance greater than $\epsilon r$ but at most $r$ from $X_h$, and let $\cS_{i_1}$ be the circle with center $X_{i_1}$ and radius $r-2s$.
At least one halfcircle of $\cS_{i_1}$ has all points at distance greater than $r$ from $X_h$, so $\ar{\cS_h\cup\cS_{i_1}}\ge (1-\epsilon^2)\pi r^2+\pi (r-2s)^2/2\ge (5/4)\pi r^2$.
Notice that, if $\cF_{i_1,i_2}$ holds for some component $\Gamma$ which contains a vertex $X_h$ such that $d(X_{i_1},X_h)\le 3r$, then we must have $d(X_{i_1},X_h)> r$ and moreover
$\cS_h\cup\cS_{i_1}$ must contain no point in $\cX\setminus\{X_{i_1},X_{i_2}\}$, which occurs with probability $(1-\ar{\cS_h\cup\cS_{i_1}})^{n-2} = O(1/n^{5/4})$.
Therefore, multiplying this by the probability that $d(X_{i_1},X_h)\le 3r$ and also taking the union bound over the $n-2$ possible choices of $h$, the contribution to $\pr(\cF_{i_1,i_2})$ due to situations of this type is $O(n(\log n/ n)/n^{5/4}) = O(\log n/n^{5/4})$.
However, we claim that this has negligible asymptotic weight in $\pr(\cF_{i_1,i_2})$.
In fact, the probability that $\cF_{i_1,i_2}$ holds for some component $\Gamma$ with all vertices at distance greater than $3r$ from $X_{i_1}$ is $\Theta(1/(n\log n))$ (in
fact we only need an upper bound and therefore we just show that this probability is $O(1/(n \log n))$.
In order to prove this last claim, we consider all the notation in the proof of Lemma~\ref{lem:EZei} for the remaining of the paragraph, and also define
$\hcS=\pi_1^{-1}(\cS)$ and $\hcY=\pi_1^{-1}(\cY)$.
Then we can repeat the same computations there but, instead of asking that all the $n-\ell$ points in $\cX\setminus\cY$ lie outside of $\cS$, we require that all the $n-\ell-2$ points in $\hcX\setminus(\hcY\cup\{\hX_{i_1},\hX_{i_2}\})$ lie outside of $\hcS\cup\hcR'_{i_1}$.
This last fact occurs with probability $\hP=(1-\vol{\hcS\cup\hcR'_{i_1}})^{n-\ell-2}$, which plays a role analogous to that of $P$.
If $X_{i_1}$ is at distance greater than $3r$ from any point in $\cY$, then
$\hcS$ and $\hcR'_{i_1}$ are disjoint. Therefore from~\eqref{eq:Sbound} we get
\begin{equation}\label{eq:Sbound_2}
\pi r^2 \left(2 + \frac{1}{6} \frac\rho{r} \right) < \vol{\hcS\cup\hcR'_{i_1}} < \frac{13\pi}{4} r^2,
\end{equation}
and an argument analogous to that leading to~\eqref{eq:Pbound} shows that
\begin{equation}\label{eq:Pbound_2}
\hP <  \left(\frac\mu n\right)^{2+\rho/(6r)}    \frac{1}{(1-13\pi r^2/4)^{\ell+1}}.
\end{equation}
Then, repeating the same computations in the proof of Lemma~\ref{lem:EZei},  but replacing $P$ with $\hP$,
proves the claim for components of type~1 of fixed size $\ell\ge2$, and this is extended to all components of 
type~1 by arguing as in Part~1 (p.L.~\ref{lem:PY}).
As a result, we conclude that $\pr(\cF_{i_1,i_2})=O(1/(n \log n))$.

Now we proceed to prove~\eqref{eq:PMi} for components of type~1. First observe that the event 
$(M_1 >0 \wedge \tK_{2,t+1}=0)$ implies that $\cE_i$ holds for some $i\in\nset$, since the only way 
for a component of type~1 to disappear within one time step is getting joined to something else. Therefore,
\[
\pr(M_1 >0 \wedge \tK_{2,t+1}=0) \le \sum_{i=1}^n \pr(\cE_i) = O\left(\frac{qn}{\log n}\right).
\]
Also notice that $(M_1 >0 \wedge B_t>0)$ implies that $\cF_{i_1,i_2}$ holds and moreover $\hX_{i_2}\in\hcQ'_{i_1}$, for some $i_1,i_2\in\nset$ ($i_1\ne i_2$). Then,
\[
\pr(M_1 >0 \wedge B_t>0) \le \sum_{i_1,i_2} \pr\big(\cF_{i_1,i_2} \wedge (\hX_{i_2}\in\hcQ'_{i_1})\big) = O\left(\frac{n^2q}{n\log n}\right) = O\left(\frac{qn}{\log n}\right).
\]
\end{pt}
\begin{pt}\label{p:dense2}
Consider all the possible components in $\nRG$ which have diameter at most $\epsilon r$ and 
size greater than  $\log n/37$. Call them components of type~2, and let $M_2$ denote their number at time $t$.

Repeat the same tessellation of \UT\ into cells as in Part~2 (p.L.~\ref{lem:PY}), and also consider the set of square boxes defined there. Given any box $b$ and $i,j\in\nset$ ($i\ne j$), we define $\cE_{b,i,j}$ to be the event that box $b$ contains more than $\log n/37-1$ points of $\cX\setminus\{X_i\}$ and moreover $\hX_i\in\hcP_j$.
Observe that each of the events $(M_2 >0 \wedge \tK_{2,t+1}=0)$ and $(M_2 >0 \wedge B_t>0)$ implies
that $\cE_{b,i,j}$ holds for some box $b$ and $i,j\in\nset$.
Then, by repeating the argument in Part~2 (p.L.~\ref{lem:PY}), but ignoring $X_i$ and also replacing $\log n/37$ with $\log n/37-1$, we deduce that
\[
\pr(M_2 >0 \wedge \tK_{2,t+1}=0) \le
O\left(\frac{1}{n^{1.1}\log n}\right)\sum_{i,j} \pr(\hX_j\in\hcP_i) = O\left(\frac{qn}{n^{0.1}\log n}\right),
\]
and the same bound applies to $\pr(M_2 >0 \wedge B_t>0)$.
\end{pt}
\begin{pt}\label{p:notembed2}
Consider all the possible components in $\nRG$ which are not embeddable but not solitary either. 
Call them components of type~4, and let $M_4$ denote their number at time $t$.

Repeat the same tessellation of \UT\ into cells as in Part~4 (p.L.~\ref{lem:PY}), and observe that
each of the events $(M_4 >0 \wedge \tK_{2,t+1}=0)$ and $(M_4 >0 \wedge B_t>0)$ 
implies that for some $i,j\in\nset$ there exists some connected union $\cS^*$ of cells 
in the tessellation with $\ar{\cS^*}\ge(11/5)\pi r^2$ such that $\cS^*\cap(\cX\setminus\{X_i\})=\emptyset$ and 
moreover $\hX_i\in\hcP_j$. Hence, from Part~4 (p.L.~\ref{lem:PY}) but replacing $\cX$ with $\cX\setminus\{X_i\}$, 
we obtain
\[
\pr(M_4 >0 \wedge \tK_{2,t+1}=0) \le
O\left(\frac{1}{n^{6/5}\log n}\right)\sum_{i,j} \pr(\hX_j\in\hcP_i) = O\left(\frac{qn}{n^{1/5}\log n}\right),
\]
and the same bound applies to $\pr(M_4 >0 \wedge B_t>0)$.
\end{pt}
\begin{pt}\label{p:medium2}
The embeddable components with diameter at least $\epsilon r$ treated in 
Part~3 (p.L.~\ref{lem:PY}) are here divided into two types.
First  consider all the possible components in $\nRG$ of diameter between $\epsilon r$ and $6\sqrt2 r$. 
Call them components of type~3a, and let $M_{3a}$ denote their number at time $t$.

We tessellate the torus \UT\ into square cells of side $\alpha r$, for some  
fixed but small enough $\alpha >0$. From Part~3 (p.L.~\ref{lem:PY}), if \RGt\ has some component 
of this type, then there exists a topologically connected union $\cS^*$ of cells with $\ar{\cS^*}\ge(1+\epsilon/6)\pi r^2$ which contains no point in $\cX$. By removing some extra cells from $\cS^*$, we can assume that the number of cells in $\cS^*$ is exactly $\lceil\frac{(1+\epsilon/6)\pi}{\alpha^2}\rceil$.
Now for each $i,j\in\nset$ and each union $\cS^*$ of 
$\lceil\frac{(1+\epsilon/6)\pi}{\alpha^2}\rceil$ cells which is topologically connected, let $\cE_{i,j,\cS^*}$ be the following event:
$\cS^*$ contains no points in $\cX\setminus\{X_i,X_j\}$, $X_j$ is at distance at least $2r$ from all the points in $\cS^*$; $\hcR'_j$ contains no points in $\hcX\setminus\{\hX_i,\hX_j\}$; and moreover $\hX_i\in\hcP_j$. Notice that if $X_j$ is at distance at least $2r$ from all the points in $\cS^*$, then $\pi_1^{-1}(\cS^*)$ and $\hcR'_j$ are disjoint. Hence, $\vol{\pi_1^{-1}(\cS^*)\cup\hcR'_j} \ge (2+\epsilon/6)\pi r^2$ and
\[
\pr(\cE_{i,j,\cS^*}) \le \left(1-\vol{\pi_1^{-1}(\cS^*)\cup\hcR'_j}\right)^{n-2} (2q)
= O\left(\frac{q}{n^{2+\epsilon/6}}\right)
\]
Similarly, let $\cF_{i,j,\cS^*}$ be the following event:
$\cS^*$ contains no points in $\cX\setminus\{X_i,X_j\}$; $X_j$ is at distance at most $2r$ from some point in $\cS^*$;  and moreover $\hX_i\in\hcP_j$. Notice that the probability that $X_j$ is at distance at most $2r$ from some point in $\cS^*$ is $O(r^2)=O(\log n/n)$. Hence,
\[
\pr(\cF_{i,j,\cS^*}) \le \left(1-\ar{\cS^*}\right)^{n-2} O\left(\frac{\log n}{n}\right) (2q)
= O\left(\frac{q\log n}{n^{2+\epsilon/6}}\right)
\]
Finally, observe that
each of the events $(M_{3a} >0 \wedge \tK_{2,t+1}=0)$ and $(M_{3a} >0 \wedge B_t>0)$ implies that 
either $\cE_{i,j,\cS^*}$ or $\cF_{i,j,\cS^*}$ hold, for some $i,j\in\nset$ and some topologically connected 
union $\cS^*$ of cells. Therefore, the probabilities of $(M_{3a} >0 \wedge \tK_{2,t+1}=0)$ and $(M_{3a} >0 \wedge B_t>0)$ are at most
\[
\sum_{i,j,\cS^*} \cE_{i,j,\cS^*} + \sum_{i,j,\cS^*} \cF_{i,j,\cS^*} =
 O\left(\frac{qn}{n^{\epsilon/6}}\right).
\]
\end{pt}
\begin{pt}\label{p:large2}
Finally consider all the possible components in $\nRG$ which are embeddable and have diameter at least $6\sqrt2 r$. 
Call them components of type~3b, and let $M_{3b}$ denote their number at time $t$.

We tessellate the torus into square cells of side $\alpha r$, for some  fixed but small enough $\alpha >0$. 
Our goal is to show that if \RGt\ has some component of type~3b, then
there exists some topologically connected union $\cS^*$ of cells with 
$\ar{\cS^*}\ge(11/5)\pi r^2$ and which does not contain any vertex in $\cX$.
Then, arguing as in Part~3', we conclude that both $\pr(M_{3b} >0 \wedge \tK_{2,t+1}=0)$ and 
$\pr(M_{3b} >0 \wedge B_t>0)$ are $O\left(qn/(n^{1/5}\log n)\right)$.
We now proceed to prove the claim on the union of cells $\cS^*$.
Given a component $\Gamma$ of type~3b in $\RGt$, let $\cS'$, $i_{\mathsf T}$ and $i_{\mathsf B}$ be defined as 
in Part~3 (p.L.~\ref{lem:PY}).
Then, by repeating the same argument in there (but replacing $\epsilon r$ with $6\sqrt2 r$), we can assume w.l.o.g. that the vertical distance between $X_{i_{\mathsf T}}$ and $X_{i_{\mathsf B}}$ is at least $6 r$, and claim that the upper halfcircle with center $X_{i_{\mathsf T}}$ and the lower halfcircle with center $X_{i_{\mathsf B}}$ must be disjoint and contained in $\cS'$. Now, consider the region of points in the torus \UT\ with the $y$-coordinate between that of $X_{i_{\mathsf T}}$ and $X_{i_{\mathsf B}}$, and split this region into three horizontal bands of the same width. Observe that each band has width at least $2r$ and hence must contain some vertex of $\Gamma$. For each of these bands, pick the rightmost vertex of $\Gamma$ in the band. We select the right lower quartercircle of radius $r$ centered at the vertex if the vertex is closer to the top of the band, or the right upper quartercircle otherwise. We also perform the symmetric operation and choose three more quartercircles to the left of the leftmost vertices in the three bands. All this six quartercircles together with the two halfcircles previously described are by construction mutually disjoint and contained in $\cS'$. Therefore $\ar{\cS'}\ge (5/2)\pi r^2$.
Let $\cS^*$ be the union of all the cells in the tessellation which are fully contained in $\cS'$. We loose a bit of area compared to $\cS'$. However, if $\alpha$ was chosen small enough, we can guarantee that $\cS^*$ is topologically connected and also $\ar{\cS^*}\ge(11/5)\pi r^2$. This $\alpha$ can be chosen to be the same for all components of type~3b.
\qedhere
\end{pt}
\renewcommand{\thept}{\arabic{pt}}
\end{proof}
%
%
Now we can characterize the connectivity of \RGT\ at two consecutive steps.
We denote by $\cC_t$ the event that $\RGt$ is connected, and by
$\cD_t=\overline{\cC_t}$ the event that $\RGt$ is disconnected.
%
%
%
\begin{cor}\label{cor:2step}
Assume that $\mu=\Theta(1)$. Then,
\begin{align*}
\pr(\cC_t\wedge\cD_{t+1}) &\sim e^{-\mu}(1-e^{-\ex B}),     & \pr(\cD_t\wedge\cC_{t+1}) &\sim e^{-\mu}(1-e^{-\ex B})
\\
\pr(\cC_t\wedge\cC_{t+1}) &\sim e^{-\mu}e^{-\ex B},     &\pr(\cD_t\wedge\cD_{t+1}) &\sim 1 - 2e^{-\mu} + e^{-\mu} e^{-\ex B}
\end{align*}
\end{cor}
\begin{proof}
First observe that $K_{1,t}=S_t+D_t$ and $K_{1,t+1} = S_t+B_t.$ Therefore we have
\[
\pr(K_{1,t}=0\wedge K_{1,t+1}>0) = \pr(S_t=0\wedge D_t=0\wedge B_t>0),
\]
and by Proposition~\ref{prop:BDS} we get
\begin{equation} \label{eq:PX0Xg0}
\pr(K_{1,t}=0\wedge K_{1,t+1}>0) \sim e^{-\ex S-\ex D}(1-e^{-\ex B}) \sim e^{-\mu}(1-e^{-\ex B}).
\end{equation}
We want to connect this probability with $\pr(\cC_t\wedge\cD_{t+1})$. In fact, by partitioning 
$(K_{1,t}=0\wedge K_{1,t+1}>0)$ and $(\cC_t\wedge\cD_{t+1})$ into disjoint events, we obtain
\begin{gather*}
\pr(K_{1,t}=0\wedge K_{1,t+1}>0) = \pr(\cC_t\wedge K_{1,t+1}>0) + \pr(\cD_t\wedge K_{1,t}=0\wedge K_{1,t+1}>0),
\\
\pr(\cC_t\wedge\cD_{t+1}) = \pr(\cC_t\wedge K_{1,t+1}>0) + \pr(\cC_t\wedge\cD_{t+1}\wedge K_{1,t+1}=0),
\end{gather*}
and thus we can write
\begin{equation}\label{eq:PCD}
\pr(\cC_t\wedge\cD_{t+1}) = \pr(K_{1,t}=0\wedge K_{1,t+1}>0) + P_1 - P_2,
\end{equation}
where $P_1  = \pr(\cC_t\wedge\cD_{t+1}\wedge K_{1,t+1}=0)$ and $P_2 = \pr(\cD_t\wedge K_{1,t}=0\wedge K_{1,t+1}>0).$

Now suppose that $s=o\big(1/(rn)\big)$. In that case, $\pr(K_{1,t}=0\wedge K_{1,t+1}>0)= \Theta(srn)$ (see~\eqref{eq:PX0Xg0} and Proposition~\ref{prop:BDS}).
Also observe that $\cD\wedge(X=0)$ implies that $\tX>0$. In fact, we must have at least two components of size greater than $1$, so at least one
of these must be non-solitary. Then, we have that $P_1 \le \pr(\tK_{2,t}=0\wedge\tK_{2,t+1}>0)$ and $P_2 \le \pr(\tK_{2,t}>0\wedge B_{t}>0),$ and
from Lemma~\ref{lem:no_birth_medium} we get
\begin{equation}\label{eq:P1P2}
P_1,P_2 = o\big( \pr(K_{1,t}=0\wedge K_{1,t+1}>0) \big).
\end{equation}
Otherwise if $s=\Omega\big(1/(rn)\big)$, then $\pr(K_{1,t}=0\wedge K_{1,t+1}>0)= \Theta(1)$.
In this case, we simply use the fact that $P_1\le \pr(\tK_{2,t+1}>0)=o(1)$ and $P_2\le \pr(\tK_{2,t}>0)=o(1)$ (see Theorem~\ref{thm:static2} and Lemma~\ref{lem:invariant}), and deduce that~\eqref{eq:P1P2} also holds.

Finally, the asymptotic expression of $\pr(\cC_t\wedge\cD_{t+1})$ is obtained  from~\eqref{eq:PX0Xg0}, \eqref{eq:PCD} and~\eqref{eq:P1P2}.
Moreover, by considering the time-reversed process, we deduce that $\pr(\cD_t\wedge\cC_{t+1})=\pr(\cC_t\wedge\cD_{t+1})$.
The remaing probabilities in the statement are computed from Corollary~\ref{cor:static} and Lemma~\ref{lem:invariant}, and using the fact that \begin{gather*}
\pr(\cC_t\wedge\cC_{t+1}) = \pr(\cC_t) - \pr(\cC_t\wedge\cD_{t+1}),
\\
\pr(\cD_t\wedge\cD_{t+1}) = \pr(\cD_t) - \pr(\cD_t\wedge\cC_{t+1}).\qedhere
\end{gather*}
\end{proof}
%
%
Let $\cA$ be an event in the static model \RG. We denote by $\cA_t$ the event that $\cA$ holds at time $t$. In the $\RGT$ model, we define $L_t(\cA)$ to be the number of consecutive steps
that $\cA$ holds starting at step $t$ (possibly $0$ if $A_t$ does not hold). Note that the distribution of $L_t(\cA)$ does not depend on $t$, and we will often omit the $t$ when it is understood or not relevant.
%
%
\begin{lem}\label{lem:2step}
Consider any event $\cA$ in the static model. If we have that $\ex(L(\cA)) <+\infty$ (but possibly $\ex(L(\cA))\to +\infty$ 
as $n\to +\infty$), then conditional upon $\cA_t$ but not $A_{t-1}$ we have
\[
\ex(L_t(\cA)\mid \overline{\cA_{t-1}}\wedge\cA_t) = \frac{\pr(\cA)}{\pr(\overline{\cA_{t-1}} \wedge \cA_t)},
\]
which does not depend on $t$.
\end{lem}

\begin{proof}
We have that
\[
L_{t-1} + 1[\overline{\cA_{t-1}}]L_t = 1[\cA_{t-1}] + L_t
\]
and taking expectations and using the hypothesis that $\ex(L(\cA))<+\infty$ we get
\[
\ex(1[\overline{\cA_{t-1}}]L_t(\cA)) = \pr(\cA),\quad \forall t.
\]
Using the fact that
\[
\ex(L_t(\cA)\mid \overline{\cA_{t-1}}\wedge\cA_t) = \frac{\ex(1[\overline{\cA_{t-1}}\wedge\cA_t] L_t(\cA))}{\pr(\overline{\cA_{t-1}}\wedge\cA_t)}
= \frac{\ex(1[\overline{\cA_{t-1}}] L_t(\cA))}{\pr(\overline{\cA_{t-1}}\wedge\cA_t)},
\]
the result follows.
\end{proof}
%
%
To prove that $\ex(L(\cC))<+\infty$ and $\ex(L(\cD))<+\infty$ we need the
following technical lemma.
%
%
\begin{lem}\label{lem:esp_finite}
Let $b=b(n)$ be the smallest natural number 
such that $(b-3)ms \geq 3\sqrt2/2$.
Then, there exists $p= p(n) > 0$ such that: for any 
fixed circle $\cR\subset\UT$ of radius $r/2$, any $i\in\nset$, any $t\in\ent$, and conditional upon any particular position of $X_{i,t}$ in the torus, the probability that $X_{i,t+b m}\in\cR$ is at least $p$.
\end{lem}
\begin{proof}
Fix an arbitrary position for $X_{i,t}$ and also for circle $\cR$, and call $X$ to its center.
Let $t'$ be the smallest integer such that $t'\mid m$ and $t'\ge t$ (i.e.\ $t'$ is the first time after $t$ when agent $i$ selects a new angle), and call $h=t'-t$, which naturally satisfies $0\le h<m$.
The particular point $X_{i,t}$ is irrelevant in our argument, and we restrict our attention to the position of agent $i$ at the times when it chooses a new angle (plus the final position), and call for simplicity
$Y_k = X_{i,t'+km}$ ($0\le k\le b-1$) and $Y_b = X_{i,t+b m}$. Observe that
\begin{equation}\label{eq:dYY}
d(Y_{k+1},Y_k)=ms, \quad \forall k \st 0\le k\le b-2,
\quad\text{and}\quad d(Y_b,Y_{b-1})=(m-h)s.
\end{equation}
Moreover recall that, if $\alpha_k$ 
denotes the angle in which agent $i$ moves between $Y_k$ and $Y_{k+1}$, 
then each $\alpha_k$ is selected uniformly and independently at random from the interval $[0,2\pi)$.

In order to prove the statement, we compute a lower bound on the probability of a strategy that is sufficient for agent $i$ to reach $\cR$ at time $t+b m$.
We start from an arbitrary point $Y_0\in\UT$ and build a sequence of points $Y_0,\ldots,Y_b$ satisfying~\eqref{eq:dYY} such that $d(Y_b,X)\le r/2$, by imposing some restrictions on the angles $\alpha_0,\ldots,\alpha_b$.
For the sake of simplicity in the geometrical descriptions, it is convenient  to allow $Y_0,\ldots,Y_b$ and $X$ to lie in $\real^2$ rather than into the torus \UT. Once the construction of the sequence of points is completed, we map them back to the torus by the usual projection.
Hence, we assume hereinafter that $Y_0$ and $X$ are two arbitrary points in $\real^2$ such that $d(Y_0,X)\le \sqrt2/2$ (which is the maximal distance in the torus \UT). 
For each $k$, $0\le k\le b-4$, we restrict $\alpha_k$ to be in $[\theta_k-\pi/6,\theta_k+\pi/6]$ (mod~$2\pi$), where $\theta_k$ is the angle of $\overrightarrow{Y_kX}$ with respect to the horizontal axis.
We claim that, with this choice of angle, the distance between $Y_k$ and $X$ is decreased at each step by at least $ms/3$ until it is at most $ms$. In fact by the law of cosines,
\begin{equation}\label{eq:lcos}
d(Y_{k+1},X) \le \sqrt{\big(d(Y_k,X)\big)^2+(ms)^2-\sqrt3d(Y_k,X)ms},
\end{equation}
and therefore, if $d(Y_k,X)>ms$, we can write
\begin{align}
d(Y_{k+1},X)
&\le \sqrt{\big(d(Y_k,X)\big)^2 + \Big(1+\frac23-\sqrt3\Big)(ms)^2 - \frac23d(Y_k,X)ms}
\notag\\
&\le \sqrt{\big(d(Y_k,X)\big)^2 + \frac19(ms)^2 - \frac23d(Y_k,X)ms}
\notag\\
&= d(Y_k,X)- \frac13ms.
\label{eq:dXY1}
\end{align}
Otherwise, if $d(Y_k,X)\le ms$, then from~\eqref{eq:lcos} we deduce that also
\begin{equation}\label{eq:dXY2}
d(Y_{k+1},X) \le \sqrt{(1-\sqrt3)\big(d(Y_k,X)\big)^2+(ms)^2} \le ms.
\end{equation}
Hence, we can guarantee that $d(Y_{b-3},X)\le ms$: Suppose otherwise that 
$d(Y_{b-3},X)> ms$. Then in view of~\eqref{eq:lcos}, \eqref{eq:dXY1} and~\eqref{eq:dXY2}, for all $k$ such that $0\le k\le b-4$ we also have $d(Y_k,X)> ms$, and moreover
\[
d(Y_{b-3},X) \le d(Y_{0},X) - (b-3)\frac{ms}3 \le \frac{\sqrt2}2 - (b-3)\frac{ms}3\le0,
\]
which contradicts the assumption.

Let $Z\in\real^2$ be the only point on the line containing $Y_{b-3}$ and $X$ satisfying $d(Z,X)=(m-h)s$ and such that 
$X$ lies on  $\overline{Y_{b-3}Z}$. Denote by $W$ one of the two points on the perpendicular bisector of  
$\overline{Y_{b-3}Z}$ which satisfy $d(W,Y_{b-3})=ms$. We want to set the angles $\alpha_{b-3}$, $\alpha_{b-2}$ and 
$\alpha_{b-1}$ so that $Y_{b-2}$, $Y_{b-1}$ and $Y_{b}$ are close to $W$, $Z$ and $X$, respectively.
Indeed, if
$\phi_{b-3}$, $\phi_{b-2}$ and $\phi_{b-1}$, respectively, are the angles between the horizontal axis and $\overrightarrow{Y_{b-3}W}$, $\overrightarrow{WZ}$ and $\overrightarrow{ZX}$, then by imposing that $\alpha_k\in[\phi_k-\epsilon r/(ms), \phi_k+\epsilon r/(ms)]$ (mod $2\pi$) for some small enough $\epsilon>0$, we achieve that $d(Y_b,X)\le r/2$ and thus $Y_b\in\cR$.

Therefore, the probability of choosing all the angles according to the strategy described is
$p = (1/6)^{b-3} \Theta\big((r/(ms))^3\big)$.
\end{proof}
%
%
The next lemma allows us to apply Lemma~\ref{lem:2step}.
\begin{lem}\label{lem:infty}
$
\ex(L(\cC))<+\infty \quad\text{and}\quad \ex(L(\cD))<+\infty.
$
\end{lem}
\begin{proof}

Fix one circle $\cR\subset\UT$ of radius $r/2$, and take $b$ as in the statement of Lemma~\ref{lem:esp_finite}.
Since the agents choose their angles independently from each other and in view of  Lemma~\ref{lem:esp_finite}, we have that,
conditional upon any arbitrary $\cX_t$, the probability that 
all agents end up inside $\cR$ after $b m$ steps is
\begin{equation}\label{eq:XRXp}
\pr(\cX_{t+bm}\subset\cR\mid\cX_t) \ge p^n,
\end{equation}
for some $p=p(n)>0$.
Observe that for any $t\in\ent$ the event $(\cX_{t}\subset\cR)$ implies that $\RGt$ is a clique, since all pairs of vertices in $\cX_t$ are at distance at most $r$, and thus $\RGt$ is connected. Consequently, for any $d\in\nat$, we can write
\begin{equation}\label{eq:PDDD}
\pr\bigg(\bigwedge_{k=0}^d \cD_{t+kbm}\bigg)
\le (1-p^n)
\pr\bigg(\bigwedge_{k=0}^{d-1} \cD_{t+kbm}\bigg)
\le \pr(\cD_{t}) (1-p^n)^d.
\end{equation}

Now observe that the equation
$L_t(\cD) = \sum_{k=0}^\infty 1[\cD_{t}]\cdots1[\cD_{t+k}],$
is satisfied pointwise, for every element in the probability space $(\cX_t)_{t\in\ent}$.
Therefore, by the Monotone Convergence Theorem, \eqref{eq:PDDD} and the fact that $p>0$, we conclude
\begin{align*}
\ex(L_t(\cD)) &= \sum_{k=0}^\infty \pr(\cD_{t}\wedge\cdots\wedge\cD_{t+k})
\\
&\le \sum_{d=0}^\infty bm \,\pr\bigg(\bigwedge_{k=0}^d \cD_{t+kbm}\bigg)
\\
&\le bm \, \pr(\cD_{t}) \sum_{d=0}^\infty (1-p^n)^d < +\infty.
\end{align*}

The same kind of argument applies to show that $\ex(L(\cC)) < +\infty$. In this case we fix two circles $\cR$ and $\cR'$ in \UT\ of radius $r/2$ and far apart from each other (say with centers at distance greater than $2r$).
Observe that for any $t\in\ent$ the event $\big((\cX_{t}\setminus \{X_{1,t}\})\subset\cR\big)\wedge(X_{1,t}\in\cR')$ implies that $\RGt$ is disconnected. Moreover, from Lemma~\ref{lem:esp_finite}, we obtain an analogue to~\eqref{eq:XRXp}
\begin{equation}\label{eq:XRXp2}
\pr\Big((\cX_{t+bm}\setminus \{X_{1,t+bm}\})\subset\cR\big)\wedge(X_{1,t+bm}\in\cR') \mid \cX_t \Big) \ge p^n,
\end{equation}
and the argument follows as in the previous case but replacing $\cD$ with $\cC$.
\end{proof}
We are now ready to prove our main theorem which characterizes the expected number of steps the graph
remains (dis)connected once it becomes (dis)connected.
\begin{thm}\label{thm:main}
\[
\ex(L_k(\cC)\mid \cD_{k-1}\wedge\cC_k) \sim \frac{1}{(1-e^{-\ex B})} =
\begin{cases}
\frac{\pi}{4srn} & \text{if } srn = o(1),\\
\frac{1}{(1-e^{-4srn/\pi})} & \text{if } srn = \Theta(1),\\
1 & \text{if } srn = \omega(1),
\end{cases}
\]
\[
\ex(L_k(\cD)\mid \cC_{k-1}\wedge\cD_k) \sim \frac{e^{\mu}-1}{(1-e^{-\ex B})} =
\begin{cases}
\frac{\pi(e^{\mu}-1)}{4srn} & \text{if } srn = o(1),\\
\frac{e^{\mu}-1}{(1-e^{-4srn/\pi})} & \text{if } srn = \Theta(1),\\
e^{\mu}-1 & \text{if } srn = \omega(1).
\end{cases}
\]

\end{thm}
\begin{proof}
Since by Lemma~\ref{lem:infty}, $\ex(L_k(\cC)) < +\infty$, $\ex(L_k(\cD)) < +\infty$, we can apply
the formula of Lemma~\ref{lem:2step} and the results follow by Corollary~\ref{cor:2step}.
\end{proof}

\section{Conclusion.}
In this paper we have formally introduced the dynamic
random geometric graph in order to study analytically dynamic MANETs. We
studied the expected length of the connectivity and
disconnectivity periods, taking into account different step sizes $s$ and different
lengths $m$ during which the angle remains invariant, always considering the
static connectivity threshold $r=r_c$. We believe that a similar analysis
can be performed for other values of $r$ as well.

The  {\em Random Walk} model simulates the behavior of a swarm of mobile vertices as sensors or robots, 
which move randomly
to monitor an unknown territory or to search in it. There exist other
models such as the {\em Random Way-point} model, where each vertex chooses
randomly a
fixed way-point (from a set of pre-determined way-points) and moves there, and when it arrives it 
chooses another and moves there,
and so on \cite{Camp02}. A possible line of future research is
to do a study similar to the one developed in
this paper for this way-point model.  We believe that the techniques developed in this paper
will prove very useful to carry out that  study.

\paragraph{Acknowledgment.} We thank Christos Papadimitriou for careful reading and 
many suggestions which improved the paper.

\end{document}